\documentclass[
	nofootinbib,
	preprint,
	noshowpacs,
	noshowkeys,
	floatfix,aps
]{revtex4-1}
\usepackage{amsmath,multirow,amssymb,mathtools}
\usepackage{verbatim}
\usepackage{float} 
\usepackage{tasks}              
\usepackage{amsfonts}
\usepackage{rotating}
\usepackage{dsfont}
\usepackage{xcolor}
\usepackage{subcaption}
\usepackage{xpatch} 
\makeatletter
	\xpatchcmd{\@ssect@ltx}{\@xsect}{\edef\@currentlabelname{#8}\@xsect}{}{}
	\xpatchcmd{\@sect@ltx}{\@xsect}{\edef\@currentlabelname{#8}\@xsect}{}{}
\makeatother
\usepackage[
	pdfstartview=FitBV,
	bookmarks=true,
	bookmarksopen=true,
	colorlinks =true,
 	linkbordercolor=blue,
 	linkcolor = blue,
	citecolor = blue,
	urlcolor=blue 
]{hyperref} 
\usepackage{tikz}
\usetikzlibrary{matrix,arrows,cd,positioning}
\newcommand{\ownref}[2]{\hyperref[#2]{#1~\ref*{#2}}}
\begin{document}
\title{ Dynamics of  polar polarizable rotors acted upon by unipolar electromagnetic pulses: From the sudden to the adiabatic regime }

\author{Marjan Mirahmadi}
\affiliation{
	Institute for Mathematics, Freie Universit\"{a}t Berlin \\ Arnimallee 6, D-14195 Berlin, Germany}
\author{Mallikarjun Karra}
\affiliation{
	Fritz-Haber-Institut der Max-Planck-Gesellschaft \\ Faradayweg 4-6, D-14195 Berlin, Germany}
\author{Burkhard Schmidt}
\email{burkhard.schmidt@fu-berlin.de}
\affiliation{
	Institute for Mathematics, Freie Universit\"{a}t Berlin \\ Arnimallee 6, D-14195 Berlin, Germany}
\author{Bretislav Friedrich}
\email{bretislav.friedrich@fhi-berlin.mpg.de}
\affiliation{
	Fritz-Haber-Institut der Max-Planck-Gesellschaft \\ Faradayweg 4-6, D-14195 Berlin, Germany}

\date{\today}

\begin{abstract}
We study, analytically as well as numerically, the dynamics that arises from the interaction of a polar polarizable rigid rotor with single unipolar electromagnetic pulses of varying length, $\Delta \tau$, with respect to the rotational period of the rotor, $\tau_r$. In the sudden, non-adiabatic limit, $\Delta \tau \ll \tau_r$, we derive analytic expressions for the rotor's  wavefunctions, kinetic energies, and field-free evolution of orientation and alignment. We verify the analytic results by solving the corresponding time-dependent Schr{\"o}dinger equation numerically and extend the temporal range of the interactions considered all the way to the adiabatic limit,  $\Delta \tau \gg \tau_r$, where general analytic solutions beyond the field-free case are no longer available. The effects of the orienting and aligning interactions as well as of their combination on the post-pulse populations of the rotational states are visualized as functions of the orienting and aligning kick strengths in terms of populations quilts. Quantum carpets that encapsulate the evolution of the rotational wavepackets provide the space-time portraits of the resulting dynamics. The population quilts and quantum carpets reveal that purely orienting, purely aligning, or even-break combined interactions each exhibit a sui generis dynamics. In the intermediate temporal regime, we find that the wavepackets as functions of the orienting and aligning kick strengths show resonances that correspond to diminished kinetic energies at particular values of the pulse duration.
\end{abstract}

\maketitle
\section{introduction}

Understanding the dynamics of quantum systems subject to strong time-dependent electromagnetic fields has been central to research areas ranging from molecular \cite{Photodissociation_using_alignedmol, surface_reactions, self_assembly, ali_hhg} to ultra-fast laser physics \cite{Dion2001, theoretical_fewhcp_ori} and from stochastic \cite{casati1979stochastic} to condensed-matter physics \cite{Izrailev_resonances_1980, QKR_AndersonLocalization}. In particular, the study of the effects on atoms and molecules of ultra-short (\textbf{$\ll$} picosecond) laser pulses and the kicks, whether single or multiple, they exert has matured into a broad field of research with a plethora of applications in science and technology, cf., e.g., Refs. \cite{app1, app2, app3, app4, app5, app6, app7}. 

For polar and polarizable molecules, the interaction with an external electromagnetic field gives rise to orienting and aligning terms, proportional to  $\cos\theta$ and $\cos^2\theta$, respectively, with $\theta$ the polar angle between the body-fixed electric dipole moment and the field vector. The electric dipole moment can be either permanent or induced, the former, denoted by $\mu$, reflecting the anisotropy of the electron distribution in the molecular frame and the latter, denoted by $\Delta \alpha$, the anisotropy of the molecule's polarizability. In 1999, it had been recognized \cite{JCP1999FriHer,FriHerJPCA99} that 
it is the combination of the aligning and orienting interactions that provides a versatile means to efficiently manipulate molecular rotation and to achieve a high degree of orientation of the molecular axis of essentially any polar molecule.  

The time dependence of the electromagnetic field plays a key role in determining the outcome of either the orienting or aligning interaction \cite{Ortigoso_BF_Dynamics,CCCC2001,Seideman_revival,Vrakking1996,Stapelfeldt_Seideman_Colloquium,Owschimikow2009,Owschimikow2011,Ghafur2009} or of their combination \cite{longcai_bf_dynamics_ali_ori,StapPRL2012}.  If the field is turned on and off slowly with respect to the molecule's rotational period, $\tau_r=\pi \hbar/B$, the molecule continues rotating as if no interaction with the field had taken place (adiabatic interaction). This is independent of how strong the effects of the field may have been during the time that the field was on. In contrast, if the field is turned on and off ``suddenly'' (i.e., over a time much shorter than the molecule's rotational period), then the orientation/alignment will recur indefinitely after the turn-off of the field, at times that are related to the molecule's rotational period (non-adiabatic interaction).\footnote{In practice, when the molecule collides, say, with the wall of the vacuum chamber, the recurrences cease as other processes take over.}  Hence molecules that suffered a sudden `kick' exhibit orientation/alignment even in the subsequent absence of a field.

In previous experiments and most theoretical studies, the aligning interaction was due to a polarized electromagnetic wave interacting with the anisotropic molecular polarizability, while the orienting interaction was brought about by subjecting the molecule to a superimposed electrostatic field. Current technologies make it possible to produce electromagnetic pulses that consist of only a few oscillation cycles \cite{hcp_modelling, hcp_gen_Bucksbaum, Voronin2014_HCP,unipolar_subpicos_plasma, picos_unipolar_photocond_switch,Bratman_quasiunipolar}. Moreover, the electromagnetic wave's electric field distribution over the cycles can have a bias, with a higher oscillation amplitude in one direction than in the other. Such `unipolar' pulses are a boon to manipulating molecular rotation: (a) Their short duration (typically less than a picosecond, as compared to nanosecond rotational periods of typical small molecules) ensures a sudden interaction that results in recurring orientation/alignment; (b) The pulse interacts simultaneously with both the permanent and induced electric dipole moment of the molecule,  whereby the combined interactions technique of manipulating molecular rotation is automatically implemented.

With few exceptions \cite{Rouzee2009, Daems2005}, a detailed time-dependent treatment of a molecular rotor interacting with a unipolar electromagnetic field has been restricted to either  a purely aligning or a purely orienting interaction \cite{Henriksen1999,Leibscher2004,Shu2010, Leibscher2003, Leibscher2004a, Rosca-Pruna2002a}, or a combination of the two, but with a time delay between them (e.g., Ref. \cite{Gershnabel2006}). Herein, we present results of a systematic study that explores both the non-adiabatic and adiabatic regime of concurrent aligning and orienting interactions for the case of a full-fledged (3D) linear molecular rotor, characterized by the values of its permanent electric dipole moment, polarizabilty anisotropy, and rotational constant. Although our  study is chiefly analytic concentrating on the effects of ultra-short pulses (or $\delta$-pulses), it also offers numerical solutions to the corresponding Schr{\"o}dinger equation involving finite-width pulses. We note that  quasi-analytic solutions of the time-independent Schr{\"o}dinger equation for the combined fields have been recently investigated via supersymmetric quantum mechanics  \cite{susy_lemeshko,susy_2dpendulum_bf_bs,Schmidt2014, Becker2017} and its complex-plane variant \cite{Schatz2018}.

This paper is organized as follows: In Section \ref{modeltdse}, we introduce the general Hamiltonian of a 3D linear rotor under  combined orienting and aligning interactions, the time-dependent Schr{\"o}dinger equation (TDSE), and the analytic and numerical approaches to solving the TDSE.  In Section \ref{results}, we  present successively the results of the population analysis, kinetic energy imparted to the molecular rotor, and the rotor's field-free evolution following its interaction with either $\delta$-pulses or finite-width pulses. Finally, in Section \ref{conclusions}, we provide a summary of our chief results and their comparison with past analyses as well as mention possible applications. 
\section{Model }\label{modeltdse}

We consider  a driven full-fledged (3D) linear rigid rotor with a  Hamiltonian of the form
\begin{equation}\label{H_genform}
\tilde{H} = B \mathbf{J}^2 + \tilde{V}(\theta,t) 
\end{equation}
where $B = \hbar^2/2I$  is the rotational constant (with $I$ the moment of inertia), 
\begin{equation}\label{angular_mom}
\mathbf{J}^2 = -\frac{1}{\sin\theta} \frac{\partial}{\partial \theta} \left(\sin\theta \frac{\partial}{\partial\theta} \right) - \frac{1}{\sin^2 \theta} \frac{\partial^2}{\partial\phi^2} 
\end{equation}
is the operator of the angular momentum squared, and $\theta \in [0,\pi]$ and $\phi \in [0,2\pi]$ are the polar and azimuthal angles \cite{Arfken2005}. 
For a dimensionless Hamiltonian 
\begin{equation}\label{Hdimensionless}
H \equiv \tilde{H} / B =\mathbf{J}^2 + V(\theta,\tau)
\end{equation}
and dimensionless time $\tau \equiv Bt/\hbar $,  the corresponding time-dependent Schr\"{o}dinger equation (TDSE) takes the form
\begin{equation}\label{scheqn}
 i \frac{\partial}{\partial\tau} \psi(\theta,\phi,\tau) = H \psi(\theta,\phi,\tau)
\end{equation}
where the corresponding dimensionless driving potential $V \equiv \tilde V/B$ is given by
 \begin{align}\label{V_genform}
 V(\theta,\tau) = \left\{
 \begin{tabular}{cr}
 $ -\eta(\tau)\cos\theta-\zeta(\tau)\cos^2\theta$  & $ 0 \leq \tau \leq \tau_0 $ \\
 0  & $ \tau > \tau_0 $
 \end{tabular} \right \}
 \end{align}
with $\eta(\tau)$ and $\zeta(\tau)$ dimensionless parameters characterizing the strengths of the orienting, $\cos\theta$, and aligning, $\cos^2\theta$,  interactions,  respectively. These interactions impart kicks (impulses) to the rotor, which we  characterize by dimensionless kick strength parameters\begin{align}\label{kickstrength}
 P_{\eta} = \int_{0}^{\tau_0}\eta(\tau)\mathrm{d}\tau , \quad
 P_{\zeta} = \int_{0}^{\tau_0}\zeta(\tau)\mathrm{d}\tau 
 \end{align}
These parameters allow for comparing the dynamics that results from kicks of different shapes, lengths, and strengths. 
Note that the energies obtained from Eq.~\eqref{scheqn} are expressed in terms of the rotational constant $B$ and that a rotational period amounts to $\tau_r=\pi$.

In what follows, we investigate the effects of the ultra-short pulses ($\delta$-pulses) that give rise to non-adiabatic interactions as well as finite-width pulses whose interactions exhibit a transition to adiabaticity.
 \subsection{$\delta$-pulses: Analytic solutions in the sudden-limit}\label{Non-adia}
 
In order to derive an analytic expression for the rotational wavepacket after the interaction 
with the potential $V$, we make use of the interaction representation \cite{Tannor2006,Sakurai1994} and write the total Hamiltonian as $ H(\tau) = H_0 + H_1(\tau) $  whose wavefunction, $\varphi$, is related to the Schr\"{o}dinger picture wavefunction, $\psi$, by $\varphi(\theta,\phi,\tau)  \equiv \mathrm{e}^{i\mathbf{J}^2\tau} \psi(\theta,\phi,\tau) $.
Thus the TDSE~\eqref{scheqn} becomes 
\begin{equation}\label{interaction_TDSE}
 i \frac{\partial}{\partial\tau}\varphi(\theta,\phi,\tau) = \left[ \mathrm{e}^{i\mathbf{J}^2\tau} V(\theta,\tau)  \mathrm{e}^{-i\mathbf{J}^2\tau}\right] \varphi(\theta,\phi,\tau) 
\end{equation}
and $\varphi$ evolves under the action of propagator $U_I$ given by the Dyson series
\begin{align}\label{propa_1}
U_I(\tau,0)  = \hat{T}\exp\left[-i\int_{0}^{\tau}\mathrm{e}^{i\mathbf{J}^2\tau'} V(\theta,\tau') \mathrm{e}^{-i\mathbf{J}^2\tau'} \mathrm{d}\tau'\right] 
\end{align}
with $\hat{T}$  the time-ordering operator. 

In the sudden limit, the duration of the interaction, $\Delta \tau$, is much less than the rotational period, $\tau_r$, i.e., $\Delta \tau /\tau_r \ll 1$. The potential of Eq.~\eqref{V_genform} then delivers a $\delta-$kick to the system, i.e.,  an instantaneous impulse, whose strength is characterized by the kick strengths $P_{\eta} $ and $P_{\zeta}$. 
In  the sudden-limit, the time ordering in Eq.~\eqref{propa_1} becomes irrelevant and operator $\mathrm{e}^{i\mathbf{J}^2\tau}$ can be neglected during the pulse ($0 \leqslant \tau \leqslant \tau_0$) \cite{Sakurai1994,Henriksen1999,CCCC2001,Matos-Abiague2006,Blanes2009}. Hence the propagator in the interaction picture can be written as
\begin{equation}
U_I(\tau > \tau_0,0) = \mathrm{e}^{-i\int_{0}^{\tau_0}V(\theta,\tau') \mathrm{d}\tau'}~. 
\end{equation} 
By taking $ \varphi(\theta,\phi,0) = \psi(\theta,\phi,0) $, the Schr\"{o}dinger picture wavefunction after the pulse ($\tau > \tau_0$) can be obtained from  
\begin{align}\label{time_eval_psi}
\psi(\theta,\phi,\tau)  &=   \mathrm{e}^{-i\mathbf{J}^2\tau}\mathrm{e}^{-i\int_{0}^{\tau_0}V(\theta,\tau') \mathrm{d}\tau'}\psi(\theta,\phi,0) \nonumber \\
&= \mathrm{e}^{-i\mathbf{J}^2\tau}\mathrm{e}^{i\left(P_{\eta}\cos\theta + P_{\zeta}\cos^2\theta\right)}\psi(\theta,\phi,0)
\end{align}
Eq.~\eqref{time_eval_psi} implies that a $\delta$-kick does not alter the probability density $\left| \psi\right| ^2$ of the system, leaving the post-pulse angular distribution the same as in the initial state -- except for a change of phase.

In what follows, we assume the system to be initially in one of the stationary states of its field free Hamiltonian, i.e., $\psi(\theta,\phi,0) = Y_{J_0}^{M_0}(\theta,\phi)$, with $Y_{J_0}^{M_0}(\theta,\phi)$ a spherical harmonic pertaining to initial quantum numbers $J_0$ and $M_0$.\footnote{Note that as a result of the azimuthal symmetry of Hamiltonian \eqref{scheqn}, the quantum number $M_0$ is conserved.} We also choose these eigenstates as a basis set for expanding the time-dependent wavefunction, with the result 
\begin{equation}\label{wf_expan_har_2}
\psi(\theta,\phi,\tau) = \sum_{J} \mathrm{e}^{-iE_J\tau}C_{J_0,M_0}^J Y_{J}^{M_0}(\theta,\phi)
\end{equation} 
where $E_J = J(J+1)$. Note that the expansion coefficients, $C_{J_0,M_0}^J$, are time-independent, in consequence of the fact that the time dependence in Eq.~\eqref{time_eval_psi} only arises from the $\mathrm{e}^{-i\mathbf{J}^2\tau}$ term.

The final form of the wavefunction in the sudden limit is given by  (for a detailed derivation see Appendix~\ref{app_CJ}) 
\begin{equation}\label{psi_final}
\psi(\theta,\phi,\tau) = \sum_{J=0}^\infty \mathrm{e}^{-iE_J\tau} \sum_{J^\prime=|J - J_0|}^{J+J_0} c^{J'} \sqrt{\frac{2J_0+1}{2J+1}} \left\langle J'0,J_00|J0 \right\rangle \left\langle J'0,J_0M_0|JM_0 \right\rangle Y_{J}^{M_0}(\theta,\phi) 
\end{equation}
where $\left\langle J_1M_1, J_2M_2 |J_3M_3\right\rangle $ are the Clebsch-Gordan coefficients and only $c^{J'}$ is a function of the kick strengths, cf. Eq.~\eqref{b_l}.

Throughout the remainder of this paper, we restrict ourselves to the case when the free rotor is initially in its ground state, $J_0=M_0=0$. As a result, the $C_{J_0,M_0}^J$ coefficients in Eq.~\eqref{wf_expan_har_2}  reduce to
\begin{equation}\label{psi_kicked_00}
C_{0,0}^J = c^J \sqrt{\frac{1}{2J+1}} 
\end{equation}

Further below, we will discuss the properties of these analytic sudden-limit wavefunctions and the expectation values obtained from them. But first, we  consider explicitly the analytic wavefunctions in the sudden limit for the purely orienting and purely aligning interactions. 
\subsubsection{Purely orienting interaction: $P_{\eta} > 0$ \& $P_{\zeta} = 0$}

This case, which is equivalent to assuming $\eta > 0$ and $\zeta = 0$ in Eq.~\eqref{V_genform} and hence $P_{\eta} > 0$ and $P_{\zeta} = 0$, has been considered  previously \cite{Henriksen1999,Leibscher2004,Shu2010}. By making use of the limit $\lim\limits_{P_{\zeta}\rightarrow 0} (P_{\zeta})^0 = 1$, one can obtain an explicit expression for $c^J$ from Eq.~\eqref{b_l} which, when substituted into Eq.~\eqref{Alm_2}, yields
\begin{equation}\label{coeff_00_pzeta0}
C_{0,0}^J = \sqrt{2J+1}\sum_{k = J}^{\infty} \frac{(iP_\eta)^k}{2^{J+1}}\frac{\Gamma((k-J+1)/2)}{\Gamma(k-J+1)\Gamma((k+J+3)/2)}
\end{equation}
Upon a change of variable, $z \equiv \frac{k-J}{2}$, and applying the Legendre duplication formula, Eq.~\eqref{coeff_00_pzeta0} reduces to 
\begin{align}\label{finalcoeff_00_pzeta0}
C_{0,0}^J &= i^{J}\sqrt{2J+1}\left[\sum_{z = 0}^{\infty}\sqrt{\frac{\pi}{2P_{\eta}}} \frac{(-1)^z}{z!(z+J+\frac{1}{2})!}\left(\frac{P_\eta}{2}\right)^{2z+J+\frac{1}{2}}\right] \nonumber \\
&= i^{J}\sqrt{2J+1}\mathcal{J}_{J}(P_\eta)
\end{align}
with $\mathcal{J}_{J}$ the spherical Bessel function of the first kind. Upon substituting for $C_{0,0}^J$ into Eq.~\eqref{wf_expan_har_2}, we finally obtain
\begin{equation}\label{final_psi_00_pzeta0}
\psi(\theta,\tau) = \sum_{J } \mathrm{e}^{-iE_J\tau}i^{J}\sqrt{2J+1}  \mathcal{J}_{J}(P_\eta) Y_{J}^{0}(\theta)
\end{equation}
\subsubsection{Purely aligning interaction: $P_{\zeta} > 0$ \& $P_{\eta} = 0$}
 
For $\eta=0$ and $\zeta>0$, the potential of Eq.~\eqref{V_genform} yields a purely aligning interaction, and so $P_{\zeta} > 0$ and $P_{\eta} = 0$. This is equivalent to setting $(k-\ell)=0$ in Eq.~\eqref{b_l}. However, in order for  $c^J$ to be non zero, $k+\ell+J$ must be even and so does $2k+J$, with the consequence  that the summation in Eq.~\eqref{psi_final} will only run over even $J$ values. The coefficients $C_{0,0}^J$ with $J$ even are given by  
\begin{equation}\label{psi_00_peta0}
	C_{0,0}^J =\sqrt{2J+1}\sum_{k = J}^{\infty} \frac{(iP_\zeta)^k}{k!}\frac{1}{2^{J+1}}\frac{\Gamma(2k+1)\Gamma((2k-J+1)/2)}{\Gamma(2k-J+1)\Gamma((2k+J+3)/2)} 
\end{equation}
while those for $J$ odd vanish. By a change of variable, $z \equiv k-J/2$, and the application of the Legendre duplication formula, the coefficients turn out to be  
\begin{align}\label{finalcoeff_psi_00_peta0}
C_{0,0}^J & = \frac{\sqrt{2J+1}}{2}(iP_{\zeta})^ {J/2} \sum_{z = 0}^{\infty} \frac{(iP_\zeta)^z}{z!}\frac{\Gamma(z+J/2+1/2)}{\Gamma(z+J+3/2)} \nonumber \\
& = \frac{\sqrt{2J+1}}{2}(iP_{\zeta})^{J/2}\left[\sum_{z = 0}^{\infty} \frac{(iP_\zeta)^z}{z!}\frac{(J/2+1/2)_z}{(J+3/2)_z}\right]\frac{\Gamma(J/2+1/2)}{\Gamma(J+3/2)} \nonumber \\
& = \frac{\sqrt{2J+1}}{2}  (iP_\zeta)^{J/2} \frac{\Gamma(J/2+1/2)}{\Gamma(J+3/2)}  {}_{1}F_1(J/2+1/2;J+3/2;iP_{\zeta})
\end{align}
where $_{1}F_1$ is the confluent hypergeometric function of the first kind (often called the Kummer function). Upon substituting for $C_{0,0}^J$ into Eq.~\eqref{wf_expan_har_2}, we finally obtain the wavefunction in the following analytic form,
\begin{equation}\label{final_psi_00_peta0}
\psi(\theta,\tau) = \sum\limits_{J=0,2,...} \mathrm{e}^{-iE_J\tau}\frac{\sqrt{2J+1}}{2}  (iP_\zeta)^{J/2} \frac{\Gamma(J/2+1/2)}{\Gamma(J+3/2)}  {}_{1}F_1(J/2+1/2;J+3/2;iP_{\zeta}) Y_{J}^{0}(\theta)  
\end{equation}
The above analytic limit has already been investigated in previous work (see, e.g., \cite{CCCC2001,Leibscher2004a,Leibscher2003}).

\subsection{Finite-width pulses: Numerical analysis}\label{finitepulse}

The more general case of finite-width pulses interacting with a polar and polarizable rigid rotor requires a numerical analysis. 
For the purposes of this analysis, we consider the temporal dependence of potential~\eqref{V_genform} to be a finite-width Gaussian, which results in the following temporal dependence of the orienting and aligning parameters,
\begin{align}
\eta(\tau) = \frac{\eta_0}{\sqrt{2\pi}\sigma}e^{-\frac{(\tau-\tau_0/2)^2}{2\sigma^2}}\label{gaussianeta}\\
\zeta(\tau)=\frac{\zeta_0}{2\pi \sigma^2}e^{-\frac{(\tau-\tau_0/2)^2}{\sigma^2}}\label{gaussianzeta}
\end{align}
with $\eta_0$ and $\zeta_0$ the amplitudes of the orienting and aligning interaction, respectively, and $\sigma$ a dimensionless pulse-width expressed in units of the rotational period of the rigid rotor. As a result,
\begin{equation}\label{P_ratio}
\frac{P_{\eta}}{P_{\zeta}} = \frac{2 \sqrt{\pi } \eta_0 
   \sigma ~
   \text{erf}\left(\frac{\text
   {$\tau $0}}{2\sqrt{2}\sigma
   }\right)}{\zeta_0 ~
   \text{erf}\left(\frac{\text
   {$\tau $0}}{2\sigma
   }\right)}
\end{equation}

In our simulations, $0\leq\tau\leq\tau_0$ with $\tau_0 = 100 ~\sigma$, and so the pulse is centered at  $\tau=\tau_0/2=50~\sigma$ and $P_{\eta}/P_{\zeta}\approx 3.545\sigma \eta_0/\zeta_0$. For the purposes of numerical integration, the choice of $100~\sigma$ very 
well approximates a Gaussian that goes to zero only at infinity, which renders the error functions in Eq.~\eqref{P_ratio} redundant. Since the Gaussians of Eqs.~\eqref{gaussianeta} and \eqref{gaussianzeta} maintain a constant area -- and thus the kicks corresponding to each impart a constant energy to the rotor, cf. Eq.~\eqref{kickstrength}, varying the pulse-width $\sigma$ offers a way of studying the non-adiabatic, transient, and adiabatic regimes at the same pulse energy, albeit with a correspondingly varying amplitude of the pulse.

We made use of the WavePacket software package \cite{Schmidt2017,Schmidt2018} for solving the TDSE, Eq.~\eqref{scheqn}, by expanding the wavefunction in a truncated orthonormal basis of Legendre polynomials $\mathcal{P}_{J}$, the so-called finite basis representation (FBR), and corresponding discrete variable representation (DVR), and using the Strang splitting propagator to switch back and forth between FBR and DVR when evaluating the effect of potential and kinetic energy, respectively.

All the numerical (in the ultra-short pulse limit) and analytic results presented in this paper were verified by comparing them with each other. Pulse-widths three orders of magnitude smaller than the rotational time period ($\sigma=0.001$) were found to be sufficiently short to reproduce the sudden limit. Therefore, all results pertaining to the sudden limit were taken straight from the analytic formulae.
\section{Results}\label{results}

We now make use of the wavefunctions obtained via the methods outlined above to study the population dynamics and field-free evolution of a polar and polarizable rotor subject to unipolar pulses/kicks. We restrict our discussion to a specific range of kick strengths, namely $P_{\eta}$ $\&$ $P_{\zeta}\in [0,10]$. This implies that the wavepacket typically comprises a moderate number of rotational states -- about $30$ for ultra-short pulses and fewer for longer pulses. 

\subsection{ Population Analysis }\label{pop}

\subsubsection{$\delta$-pulses}

\begin{figure}[H]
	\centering
	\includegraphics[scale=0.34]{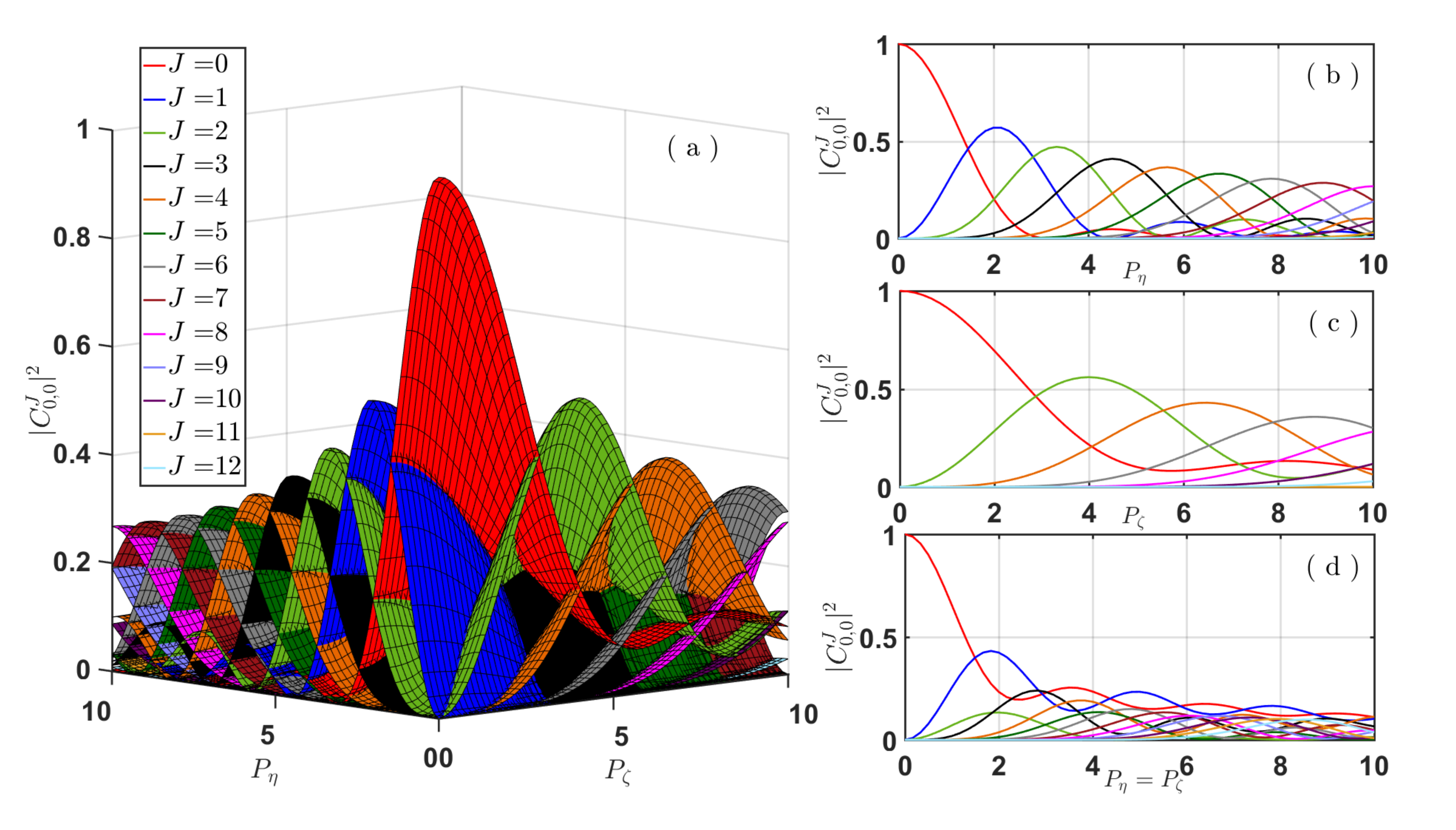}
	\caption{ \label{section} Populations, $|C_{0,0}^J|^2$, arising due to the $\delta$-kicks as derived from Eq.~\eqref{C_lm0} with $J_0=0$ for (a) different $P_{\eta} $ and $P_{\zeta}$ values, (b) different $P_\eta $ values at $P_\zeta= 0$, (c) different $P_\zeta$ values at $P_\eta = 0 $ and (d) $P_\eta = P_\zeta $.  }
\end{figure}

\begin{figure}[H]
	\centering
	\includegraphics[scale=0.4]{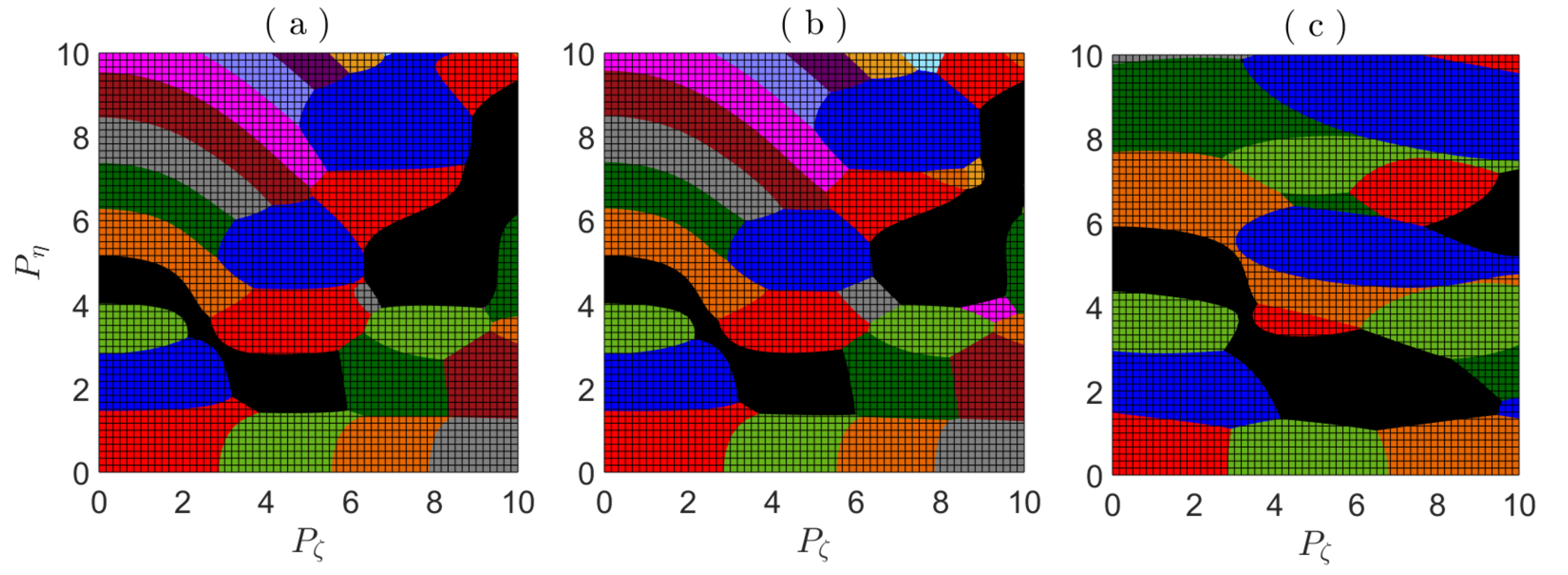}
	\caption{\label{quilt} Population quilts arising from (a) $\delta$-kicks as derived from Eq.~\eqref{C_lm0} with $J_0=0$, (b) finite length pulses with $\sigma = 0.01$ and  (c) $\sigma = 0.1$. The color coding of the $J$ state populations is the same as in Fig.~\ref{section}. }
\end{figure}

The populations (probability densities), $|C_{0,0}^J|^2$, see Eqs.~\eqref{psi_final} and \eqref{psi_kicked_00}, of different $J$-states that make up the post-kick rotational wavepacket are illustrated in Figure~\ref{section}. Note that the rotor was initially in its ground state $J_0=0$.
A top view (projection onto the $P_{\eta},P_{\zeta}$ plane) of the populations shown in Fig.~\ref{section}a gives rise to  a ``population quilt'' (Fig.~\ref{quilt}a) that displays the dominant contributions from the different rotational states to the wavepacket created by various combinations of the aligning and orienting kick strengths.
In the non-adiabatic regime, Fig.~\ref{quilt}a and b, we note three main domains -- one along the $P_{\eta}=P_{\zeta}$ diagonal, and one on either side of it. 

The effect of a purely orienting interaction can be seen in Fig.~\ref{quilt}a in the vicinity of the $P_{\eta}$ axis. In this domain, the general behaviour of the spherical Bessel function in Eq.~\eqref{finalcoeff_00_pzeta0} dominates -- i.e., with increasing $P_{\eta}$, the quantum number of the most populated rotational state increases and the selection rule $\Delta J = \pm 1$ for the rotational transitions applies.

In contrast, for a purely aligning interaction (see the part of the quilt in the vicinity of the $P_{\zeta}$ axis), the dominant contribution comes from the confluent hypergeometric function ${}_{1}F_1$ in Eq.~\eqref{finalcoeff_psi_00_peta0} and the most populated rotational states have even $J$'s following the selection rule $\Delta J = \pm 2$. Like for the purely orienting interaction, the most populated $J$ values increase with $P_{\zeta}$.

Starting from either of the two above domains, upon increasing or decreasing the lopsided ratio of the orienting and aligning kick strengths and moving closer to the diagonal where they are equal, the most populated states (with a minor exception around $P_{\eta}= P_{\zeta} = 3$) become those with $J=0$ and $J=1$ that alternate as $P_{\eta}$ and $P_{\zeta}$ increase along the diagonal,  see Fig.~\ref{quilt}a.  

\begin{figure}[H]
	\centering
	\includegraphics[scale=0.32]{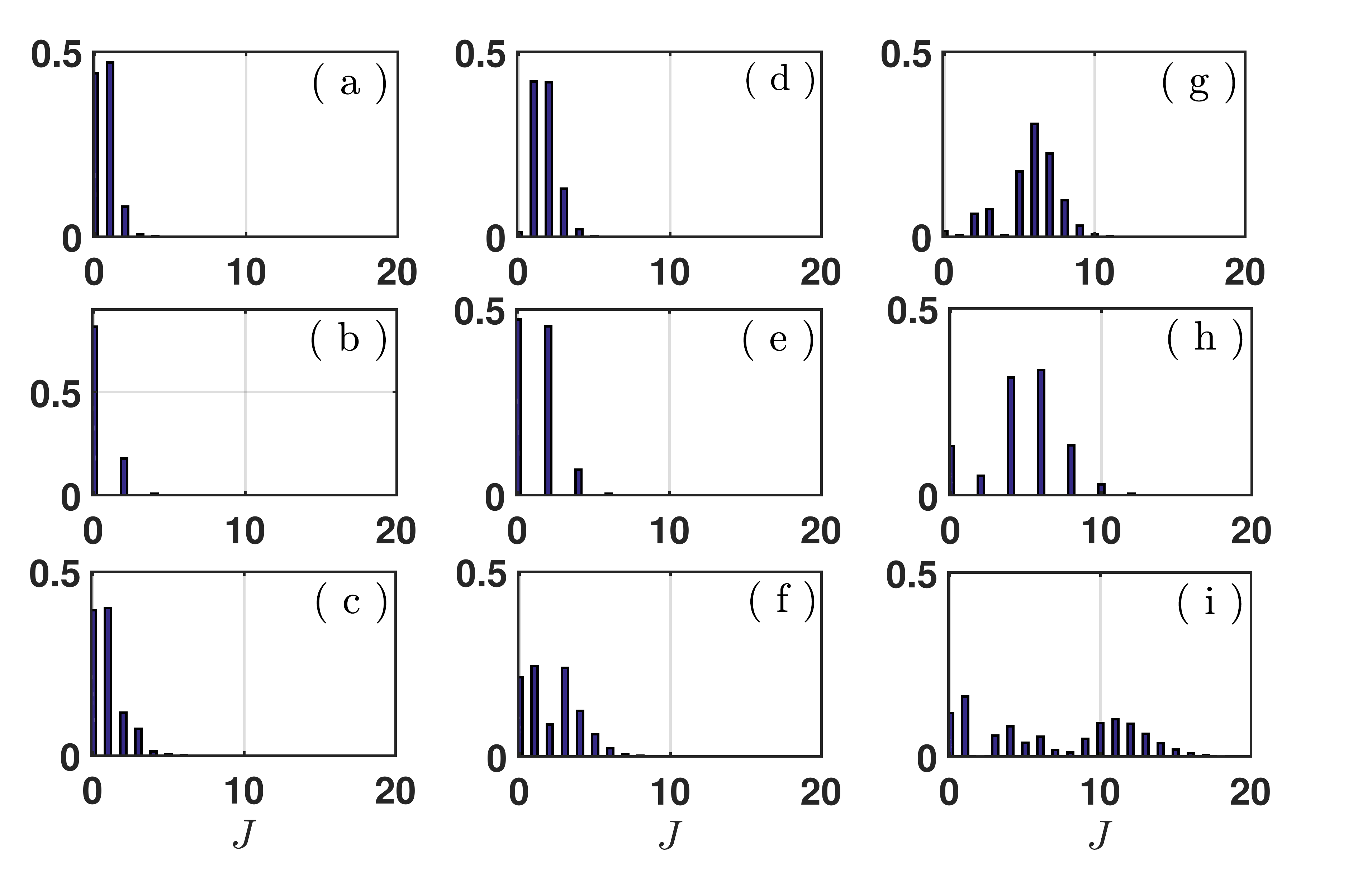}
	\caption{\label{coeff} Post-pulse populations $|C_{0,0}^J|^2$ of $J$ states arising form $\delta$-kicks with $P_{\eta} = P, P_{\zeta} = 0$ (panels a, d, g), $P_{\eta} = 0, P_{\zeta} = P$ (panels b, e, h), and $P_{\eta} = P_{\zeta} = P$ (panels c, f, i), where $ P = 1.5 $ in panels (a-c), $ P = 2.8$ in panels (d-f), and $ P = 8 $ in panels (g-i).}
\end{figure}

Fig.~\ref{coeff} illustrates the population dynamics due to a $\delta$-pulse for some particular choices of points in the $(P_{\eta},P_{\zeta})$ plane. This is complemented by planar projections of the populations displayed in Fig.~\ref{section} and shown in Figs. ~\ref{section} b, c, d.  We see that close to $P_{\eta} = 1.5$ ($P_{\zeta} = 0$), the wavepacket is essentially an equally-weighted superposition of the two lowest states, $J =0$ and $J =1$  (Figs.~\ref{section}b and \ref{coeff}a). Furthermore, at $P_{\zeta} = P_{\eta} \approx 1.5$, these two states remain the two highest-populated states, each contributing about $2/5$ to the overall population (Fig.~\ref{section}d and \ref{coeff}c). 
This behaviour has a counterpart for $ P_{\zeta} = 0 $  and $ P_{\zeta} \approx 2.8 $, with the dominant $J = 0$ and $ J = 2$ states  contributing almost equally, see Figs.~\ref{section}c and \ref{coeff}e. On the diagonal at $P \approx 2.8$ (Fig.~\ref{section}d), three states ($J=0,1,3$) possess nearly the same populations, albeit with $J = 3$ slightly higher than the other two.

The $P = 8 $ case shows that in the strong kick-strengh regime the number of fractional revivals increases and that orientation and alignment undergo rapid changes compared with weaker pulses as reflected in the highly  structured  quantum carpets (see Subsection \ref{S-T}). Also the quantum-classical correspondence becomes manifest (see Subsection \ref{kres}).

\subsubsection{Finite-width pulses}

For the finite-width pulses, we studied the effect of the pulse-width by varying $\sigma$ between $0.001$ and $1$, thereby imparting the same kick-energies over longer time spans. Whereas for $\sigma=0.001$ we reproduce much of the behaviour found analytically for the $\delta$-pulses, at $\sigma=1$ the dynamics begins to slowly approach the adiabatic limit, i.e., the system is left without any excitation at the end of the pulse and the initial rotational state ($J=0$) remains the most populated state throughout the domain of the kick strengths considered.

In particular, for $\sigma=0.01$ (Fig.~\ref{quilt}b), the populations closely resemble those obtained in the sudden regime, except for (a) an enlargement of some existing regions of higher populations (namely, $J=4$ around $P_{\eta} = P_{\zeta} \approx 10$, $J=6$ around $P_{\eta} \approx 4$, $P_{\zeta} \approx 6$ and $J=12$ around $P_{\eta} \approx 10$, $P_{\zeta} \approx 6$, and (b) the unexpected surfacing of hillocks due to higher populated states, namely $J=4$ around $P_{\eta} \approx 7$, $P_{\zeta} \approx 8.3$, $J=8$ around $P_{\eta} \approx 4$, $P_{\zeta} \approx 9$ and $J=11$ around $P_{\eta} \approx 7$, $P_{\zeta} \approx 8.8$.

At a ten-fold pulse-width, $\sigma=0.1$, a transition to adiabaticity becomes manifest, with about half as many dominant states populated as in the non-adiabatic, sudden regime. Moreover, a washing out of the three domains observed in the sudden limit takes place, as evident from Fig.~\ref{quilt}c. Indeed, the corresponding population quilt looks  like a ``zoomed-in'' lower left corner of the quilts shown in Figs.~\ref{quilt}a,~b.

\subsection{Kinetic energy}\label{kres}

\subsubsection{$\delta$-pulses: the classical-quantum correspondence}

The kinetic energy imparted to the rotor, i.e., the expectation value of its angular momentum squared at the end of the kick, 
\begin{equation}\label{kinetic}
\left\langle \mathbf{J}^2 \right\rangle=\sum_{J} J(J+1) \left|C_{0,0}^J\right|^2 
\end{equation}
can be obtained by evaluating $\left\langle \mathbf{J}^2 \right\rangle$  from Eqs.~\eqref{angular_mom} and \eqref{time_eval_psi} at time $\tau_0$, with the result,
\begin{align}\label{kin_tauplus}
\left\langle \psi(\tau_0) \left| \mathbf{J}^2 \right| \psi(\tau_0) \right\rangle &= 
\int_0^{2\pi}\mathrm{d}\phi \int_0^{\pi} \sin\theta \mathrm{d}\theta Y_0^0 \mathrm{e}^{-i\left(P_{\eta}\cos\theta + P_{\zeta}\cos^2\theta\right)} \mathbf{J^2} \mathrm{e}^{i\left(P_{\eta}\cos\theta + P_{\zeta}\cos^2\theta\right)} Y_0^0 \nonumber \\
&= \frac{i}{2} \int_0^{\pi} \sin\theta \mathrm{d}\theta \left[ 2P_{\zeta}(2\cos^2\theta-1)-iP_{\eta}^2\sin^2\theta-4iP_{\zeta}^2\cos^2\theta\sin^2\theta  \right]  \nonumber \\
& =  \frac{2}{3}P_{\eta}^2 + \frac{8}{15} P_{\zeta}^2 
\end{align}
All other contributions to the integral vanish by symmetry. Thus the dependence of the kinetic energy imparted by a $\delta$-pulse on the kick strengths is monotonous -- a surprisingly simple result, especially when compared with the rich -- and partly counterintuitive -- structure of the population quilts.

Interestingly, Eq.~\eqref{kin_tauplus} has a classical counterpart which is isomorphic with it. The Lagrange equation of motion of a rotor in potential $V(\theta,\tau)$ given by Eq.~\eqref{V_genform} for variable $\theta$ reads
\begin{align}\label{eqmotion}
\frac{\mathrm{d}}{\mathrm{d}\tau}\left( \frac{1}{2}\frac{\mathrm{d}\theta}{\mathrm{d}\tau}\right)  &= - \frac{\partial V}{\partial\theta} \nonumber \\
&=-\eta(\tau)\sin\theta-\zeta(\tau)\sin 2\theta 
\end{align}
with the dimensionless time $\tau \equiv t/2I$, potential $2I\tilde{V}\equiv V$, and kinetic energy $\frac{1}{4}\left(\frac{\mathrm{d}\theta}{\mathrm{d}\tau}\right)^2$ in the units of $1/2I$. Accordingly, for a rotor with zero initial angular velocity and an initial coordinate $\theta_i$ (i.e., before the pulse), integration of  Eq.~\eqref{eqmotion} over the pulse duration yields an angular velocity $\frac{\mathrm{d}\theta_i}{\mathrm{d}\tau} = -2 \left( P_{\eta}\sin\theta_i + P_{\zeta}\sin 2\theta_i \right) $ after the pulse. As a result, the mean value of the dimensionless classical kinetic energy over different initial angles then becomes
\begin{align}\label{classicalkin}
\left\langle \mathbf{J}_{\mathrm{cl}}^2\right\rangle &= \frac{1}{2} \int_0^{\pi} \sin\theta_i \mathrm{d}\theta_i \left[ \frac{1}{4}\left(\frac{\mathrm{d}\theta_i}{\mathrm{d}\tau}\right)^2 \right] \nonumber \\
&= \frac{1}{2}\int_0^{\pi} \sin\theta_i \mathrm{d}\theta_i\left[P_{\eta}\sin\theta_i + P_{\zeta}\sin 2\theta_i \right]^2 \nonumber \\
& =  \frac{2}{3}P_{\eta}^2 + \frac{8}{15} P_{\zeta}^2 
\end{align}  
which is isomorphic with Eq.~\eqref{kin_tauplus}, as noted.

\subsubsection{Finite-width pulses: $\sigma$-dependent resonances}

During the pulse, the kinetic energy of the rotor increases, reaches a maximum, and then decreases to a constant, post-pulse value. This post-pulse value decreases with increasing pulse-width as the adiabatic limit is approached. However, the dependence of the post-pulse kinetic energy of the rotor on the pulse-width  $\sigma$ exhibits a previously overlooked behaviour.

\begin{figure}[H]
	\centering
	\includegraphics[scale=0.31]{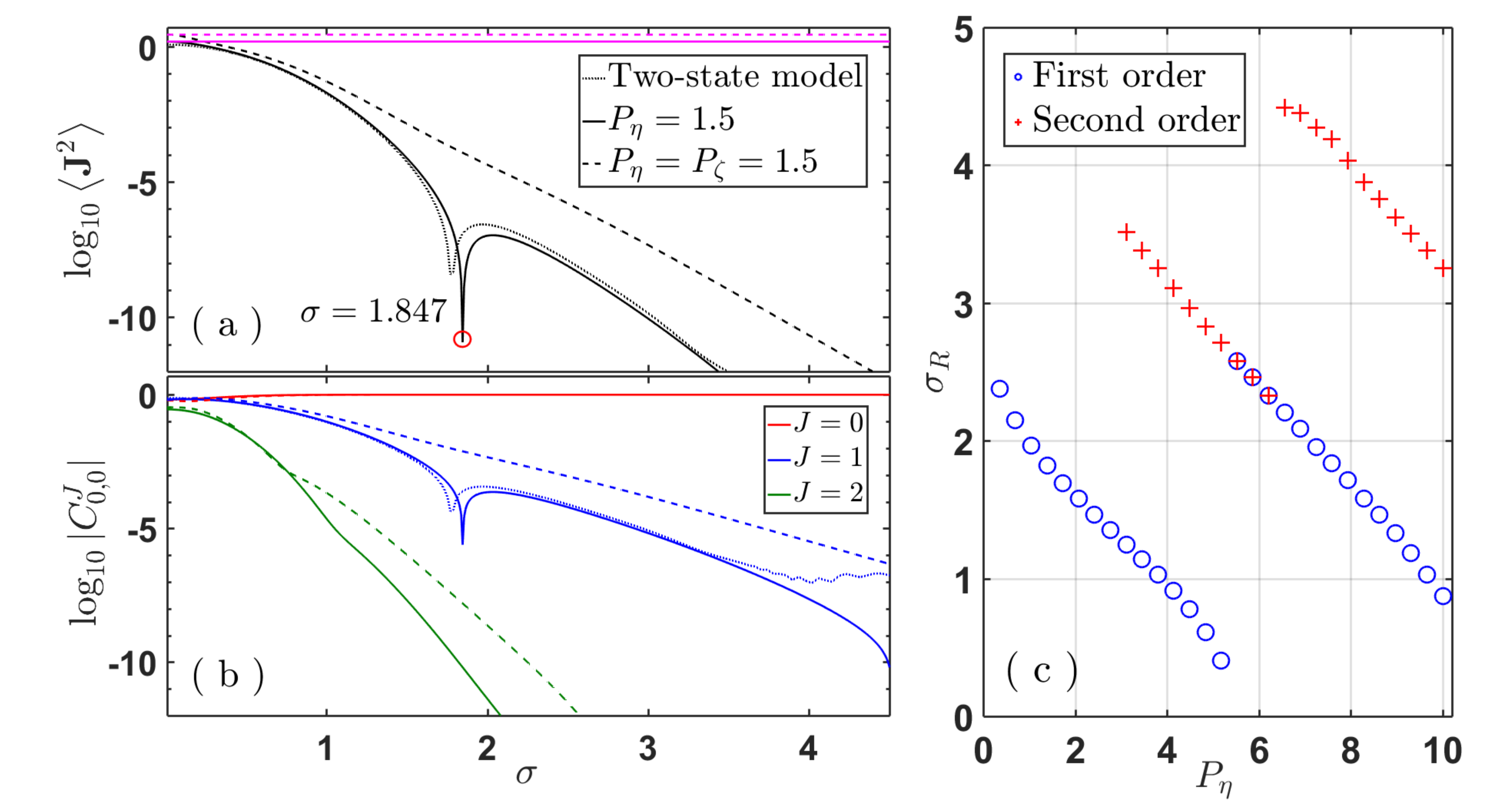}
	\caption{(a) Semi-logarithmic representation of the kinetic energy imparted by
pulses of varying widths, $\sigma$, as obtained from the two-level model (black dotted line) or a numerical calculation for either a purely orienting interaction with $P_{\eta} = 1.5$ (solid black line) or a combined interaction with $P_{\eta} = P_{\zeta} = 1.5$ (black dashed line). The two horizontal lines in magenta denote the energies in the sudden limit, Eq.~\eqref{kin_tauplus}. (b) Numerically calculated coefficients of the rotational levels $J = 0$ (red line), $J = 1$ (blue line), and $J = 2$ (green line) for $P_{\eta} = 1.5$ (solid lines) and $P_{\eta} = P_{\zeta} = 1.5$ (dashed lines) compared with the coefficients of the $J = 0$ (dashed red line, coinciding with the solid red line) and $J = 1$ (dashed blue line) levels obtained from the two-level model. We observe a robust resonance at $\sigma_R = 1.847$ that is close to that predicted by the two-level model. (c) Scatter plot showing the positions, $\sigma_R$, of first- and second-order resonances as a function of $P_{\eta}$ for purely orienting pulses.}
	\label{kineticsigma}
\end{figure}

In the semi-logarithmic plot in panel (a) of Fig.~\ref{kineticsigma}, which shows the post-pulse kinetic energy as function of the pulse duration $\sigma$ for $P_{\eta} = 1.5 \mathrm{ \ and \ } P_{\eta} = P_{\zeta} = 1.5$, three separate regimes can be distinguished: (i) The non-adiabatic regime, on the far left where $\sigma\rightarrow0$. The corresponding kinetic energy is nearly constant, given by Eq.~\eqref{kin_tauplus} and represented by the  horizontal magenta lines. (ii) The transition regime at medium $\sigma$ ($1.5 \lesssim \sigma \lesssim 3$) where a conspicuously sharp resonance occurring at $\sigma = 1.847$ (about $0.6$ times the rotational time-period) effects a sudden drop in kinetic energy by nearly three orders of magnitude. Thus, we see the possibility of a rotational arrest much before the adiabatic regime is reached. Since the kinetic energy comprises a sum of squares of coefficients $C_{0,0}^J$, cf.  Eq.~\eqref{kinetic}, the resonances occur whenever the coefficients contributing to the wavepacket vanish (see Fig.~\ref{kineticsigma}b). (iii) The adiabatic regime, $\sigma >3$, where the kinetic energy drops off by ten orders of magnitude compare to the ultra-short pulse regime (see Fig.~\ref{kineticsigma}a). The combined fields pulse behaves differently in the medium $\sigma$ regime, with the resonance(s) (observed for higher kick strengths) pushed rightwards, while closely mimicking the purely orienting pulse in the other two regimes.

In order to better understand the origin of the resonances, we set up a two-level model  with the free-rotor ground state, $Y_{0}^0$, being the initial state (before the arrival of the pulse) and the superposition of the ground and first excited rotational states, $Y_{0}^0$ and $Y_{1}^0$, being the final state (at the end of the pulse) \cite{Tannor2006}. 
The two-level Hamiltonian matrix in the basis of the $Y_{0}^0$ and $Y_{1}^0$ free-rotor states reads,  
\begin{eqnarray}\label{twolevelH}
H(\tau) = \begin{pmatrix}
0 & V(\theta,\tau) \\
V(\theta,\tau) & 2
\end{pmatrix} = \begin{pmatrix}
0 & \frac{-\eta(\tau)}{\sqrt{3}} \\
\frac{-\eta(\tau)}{\sqrt{3}} & 2
\end{pmatrix}
\end{eqnarray}
where we made use of the matrix element  $\left\langle Y_{0}^0|\cos\theta|Y_{1}^0\right\rangle = 1/\sqrt{3}$. We note that in the medium-$\sigma$ domain, the two-level model is only approximate -- see the non-negligible coefficient of the $J = 2$ state in the lower panel of Fig.~\ref{kineticsigma} -- but becomes accurate for higher $\sigma$ values and even morphs into an essentially one-level system in the adiabatic regime. 
Since the off-diagonal matrix element $\left\langle Y_{0}^0|\cos^2\theta|Y_{1}^0\right\rangle$ pertaining to the aligning interaction vanishes, the two-level model only captures the orienting and combined interactions but not the purely aligning one. We note that the two-level model describes adequately the effect of the purely orienting pulse over the whole $\sigma$ domain -- even in the non-adiabatic limit; however, this is only because of our particular choice of kick strengths, while for other choices, a two- or three-level approximation holds only in the medium $\sigma$ regime.

In the interaction picture, cf. Eq.~\eqref{interaction_TDSE}, the matrix form of the TDSE  within the two-level model becomes
\begin{align}
i \frac{\mathrm{d}}{\mathrm{d}\tau} \begin{pmatrix} C_{0,0}^0(\tau)\\
C_{0, 0}^{1}(\tau) \end{pmatrix} = \begin{pmatrix} 1 & 0 \\
0 &	e^{2i\tau} \end{pmatrix} \begin{pmatrix}
0 & \frac{-\eta(\tau)}{\sqrt{3}} \\
\frac{-\eta(\tau)}{\sqrt{3}} & 2
\end{pmatrix}  \begin{pmatrix} 1 & 0 \\
0 &	e^{-2i\tau} \end{pmatrix} \begin{pmatrix} C_{0, 0}^0(\tau)\\
C_{0, 0}^{1}(\tau) \end{pmatrix}
\end{align}
which results in the following set of coupled differential equations for the coefficients $C_{0, 0}^0$ and $C_{0, 0}^1$,
\begin{align}
	\frac{\mathrm{d}}{\mathrm{d}\tau}C_{0, 0}^0(\tau) &= i e^{-2i\tau}\bigg(\frac{\eta(\tau)}{\sqrt{3}}\bigg)C_{0, 0}^{1}(\tau), \nonumber \\
	\frac{\mathrm{d}}{\mathrm{d}\tau}C_{0, 0}^1(\tau) &= i e^{2i\tau}\bigg(\frac{\eta(\tau)}{\sqrt{3}}\bigg)C_{0, 0}^{0}(\tau)
\end{align}
By solving the above two coupled equations numerically, we obtain the coefficients of the ground and excited states as a function of the dimensionless time $\tau$, and hence, the post-pulse kinetic energy, $ \left\langle \mathbf{J}^2\right\rangle =2|C_{0, 0}^1(\tau_0)|^2$). This is shown in panel (a) of Fig.~\ref{kineticsigma} by the black dotted curve (pertaining to $P_{\eta}=1.5$). Furthermore, by successively
solving the said coupled equations numerically, we obtain a scatter plot shown in panel (c) of Fig. ~\ref{kineticsigma}, which displays the positions, $\sigma_R$, 
of various first- and second-order resonances as a function of $P_{\eta}$. It is found that the various orders of resonances oscillate with the same frequency in $P_{\eta}$ (of about $5.5$) -- a behaviour that may be described as  a ``forward cycling oscillation'' in  $\sigma$. Higher orders are washed out as one approaches the adiabatic limit, but become visible again with increasing  kick strength $P_{\eta}$. The behaviour of these resonances for higher kick strengths as well as their connection to orientation and alignment will be treated elsewhere. 

\subsection{Space-time portraits of the probability densities of rotational wavepackets created by $\delta$-pulses}\label{S-T}

In this subsection as well as the following one, we analyze the dynamics of the polar polarizable rotor subject to $\delta$-pulses based on the analytic theory developed in Subsection \ref{Non-adia}. The characteristic phenomena that such an interaction brings about, such as wavepacket revivals, (anti-)orientation and (anti-)alignment are captured by the so-called ``quantum carpets,'' i.e., contour plots of the probability density $|\psi(\theta,\tau)|^2$ in the $\theta,\tau$ plane \cite{Grossmann1997,Stifter1997}. Note that we use the common definition of alignment, $\theta \approx 0$ or $ \pi$, anti-alignment,  $\theta \approx \pi/2$, as well as orientation, $\theta \approx 0$ and anti-orientation, $\theta \approx \pi$. behaviour

The wavepacket, Eq.~\eqref{wf_expan_har_2}, can be recast in terms of the modulus and phase  of the expansion coefficients $C_{0,0}^J$,
\begin{equation}\label{wfphasemodulus}
\psi(\theta,\tau) = \sum_{J} \left|C_{0,0}^J \right| Y_{J}^{0}(\theta)  e^{i(\gamma_J-E_J\tau)} 
\end{equation}
with $e^{i\gamma_J} = C_{0,0}^J /\left|C_{0,0}^J \right| $ and so the probability density can be written as 
\begin{align}\label{prphasemodulus}
|\psi(\theta,\tau)|^2 = \sum_{J} \left|C_{0,0}^J \right|^2 \left[Y_{J}^{0}(\theta)\right]^2 
+ 2 \sum_{J > J'} \left|C_{0,0}^J\right| |C_{0,0}^{J'}| Y_{J}^{0}(\theta) Y_{J'}^{0}(\theta) \cos\left(\Delta\gamma_{JJ'}-\Delta E_{JJ^\prime}\tau\right)
\end{align}
with $\Delta\gamma_{JJ^\prime}\equiv \gamma_J-\gamma_{J'} ,\Delta E_{JJ^\prime}\equiv  E_J- E_{J'} $. Thus the probability density consists of a constant ``population'' term and an oscillatory ``coherence'' term, the latter comprised of  Fourier terms with different frequencies $\Delta E_{JJ^\prime}$ but with a constant phase shift $\Delta\gamma_{JJ^\prime}/\Delta E_{JJ^\prime}$ for each frequency.

Since $E_J=J(J+1) $ is an even integer, the wavefunction of Eq.~\eqref{wfphasemodulus} is identical at the end of the pulse to its initial value at integer multiples of the revival time $\tau_{\mathrm{rev}}=\pi$, 
\begin{align}\label{fullrev}
\psi(\theta,\tau+\pi) = \psi(\theta,\tau) ~.
\end{align}
Moreover, for  $N = J-J'$,  $\Delta E_{JJ^\prime}= 2J'N+N(N+1)$ is an even integer too, and so the coherence term in Eq.~\eqref{prphasemodulus} is a $\pi$-periodic function represented by a Fourier sum of terms with periods
\begin{align}\label{fractionalrev}
\frac{2\pi}{\Delta E_{JJ^\prime}} = \frac{\tau_{\mathrm{rev}}}{J'N+N(N+1)/2}
\end{align}
each of which is an even or odd integral fraction of the revival time. The presence of  even fractions explains the absence of a mirror revival (reformation of the initial wavepacket with a changed sign in a mirrored location about $\theta = \pi/2$  \cite{Robinett2004}) at half of the revival time $\tau=\pi/2$. 

Furthermore, at half of the revival time, because of different values of the phase $e^{-iE_J\pi/2}$ for $J(\bmod{4})\in\left\lbrace0,1,2,3\right\rbrace $, the wavepacket of Eq.~\eqref{wfphasemodulus} can be split into two terms 
\begin{align}\label{halfmods} 
\psi(\theta,\tau+\pi/2) = \sum_{J(\bmod{4}) \in \left\lbrace 0, 3\right\rbrace } \left|C_{0,0}^J \right| Y_{J}^{0}(\theta)  e^{i(\gamma_J-E_J\tau)} - \sum_{J(\bmod{4})\in\left\lbrace 1, 2\right\rbrace } \left|C_{0,0}^J \right| Y_{J}^{0}(\theta)  e^{i(\gamma_J-E_J\tau)}
\end{align}
representing a fractional revival. Wavepackets with only $J(\bmod{4}) \in \left\lbrace 0, 3\right\rbrace $ as the main populated states will closely resemble the initial state after half of the revival time. Similarly, wavepackets with only $J(\bmod{4}) \in \left\lbrace 1, 2\right\rbrace $ as the main populated states, will also resemble the initial state but with the opposite sign. 

\subsubsection{Purely orienting interaction} 

The quantum carpets for the purely orienting interaction are shown in panels (a), (d), and (g) of Figure \ref{carpets}. A conspicuous feature they exhibit  is the inversion symmetry of the probability density $|\psi(\theta,\tau)|^2$ with respect to the point $\theta=\tau=\pi/2$. This inversion symmetry has its origin in the equality
\begin{equation}\label{zeta0half}
\psi\left(\theta,\tau\right) = \psi^*\left(\pi-\theta,\pi-\tau\right)
\end{equation}  
which follows from the wavefunction of Eq.~\eqref{final_psi_00_pzeta0}, where $e^{i\gamma_J}=i^J$, and the parity transformation $Y_{J}^M (\pi-\theta,\pi+\phi) = (-1)^{J} Y_{J}^M (\theta,\phi) $.
Eq.~\eqref{zeta0half} leads to
\begin{equation}
\psi\left(\theta,\frac{\pi}{2}\right) = \psi^*\left(\pi-\theta,\frac{\pi}{2}\right) 
\end{equation} 
which implies the splitting of the wavepacket into two separate packets at $\tau=\tau_{\mathrm{rev}}/2=\pi/2$, see panels (a), (d), and (g) of Fig.~\ref{carpets}.

\subsubsection{Purely aligning interaction}\label{S-T_Pzeta} 
Analogous results to those above are not available for the purely aligning interaction, due to the complicated phase and modulus terms arising in the corresponding wavefuntion, Eq.~\eqref{final_psi_00_peta0}. However, we can glean some general features of the purely aligning interaction from the quantum carpets shown in panels (b), (e), and (h) of Fig.~\ref{carpets}. 

Shortly after $\tau= \tau_{\mathrm{rev}}/2=\pi/2$, the probability densities have  almost isotropic angular distributions, which are reflections of the initial angular distribution.
Such isotropic distributions are mostly found at weak kick-strengths, see panel (b) of Fig.~\ref{carpets}. With increasing $P_{\zeta}$, the probability densities become increasingly anisotropic.

We also note that whereas aligned and anti-aligned distributions appear, respectively,  before and after $\tau= \tau_{\mathrm{rev}}/2$, this order is reversed at the full revival time $\tau= \tau_{\mathrm{rev}}$ (more about this in Section~\ref{oriali}).

Furthermore, at $\tau= \tau_{\mathrm{rev}}/4$ and $\tau= 3\tau_{\mathrm{rev}}/4$, partial anti-alignment and alignment occurs, respectively. Applying a phase equivalent to $e^{-\frac{i\pi}{4}}$, Eq.~\eqref{halfmods} yields
\begin{align}\label{pi3414}
\psi(\theta,\tau+\frac{\pi}{2}\mp\frac{\pi}{4}) &= \sum_{J(\bmod{8}) = 0} \left|C_{0,0}^J \right| Y_{J}^{0}(\theta)  e^{i(\gamma_J-E_J\tau)} - \sum_{J(\bmod{8}) = 4} \left|C_{0,0}^J \right| Y_{J}^{0}(\theta)  e^{i(\gamma_J-E_J\tau)} \nonumber \\
 &\pm i \left[ \sum_{J(\bmod{8}) = 2} \left|C_{0,0}^J \right| Y_{J}^{0}(\theta)  e^{i(\gamma_J-E_J\tau)} - \sum_{J(\bmod{8}) = 6} \left|C_{0,0}^J \right| Y_{J}^{0}(\theta)  e^{i(\gamma_J-E_J\tau)} \right] ~.
\end{align}
For $\tau=0$  the sign of coefficients with $J(\bmod{4})=0$ does not change, in contrast to the sign of coefficients with $J(\bmod{4})=2$. 
Even though an explicit formula is elusive, using the following relations for spherical harmonics (with $J$ even)   
\begin{align}
Y_J^0 (\theta) &= \sum_{n=0}^{J} a_{n}^J \cos(n\theta) \nonumber \\
Y_J^0 \left(\frac{\pi}{2}-\theta \right) &= \sum_{n=0}^{J} (-i)^{n} a_{n}^J \cos(n\theta)  
\end{align}
where $a_{n}^J \in \mathbb{R^+}$, and noting that $(-i)^{J(\bmod{4})=0}=1 ,(-i)^{J(\bmod{4})=2}=-1$, we can qualitatively interpret the change of sign in Eq.~\eqref{pi3414} as a $\pi/2$-rotation and infer that 
\begin{align}\label{re}
\psi \left(\theta,\frac{\pi}{4}\right) \sim \psi\left(\frac{\pi}{2}-\theta,\frac{3\pi}{4}\right) 
\end{align} 

\subsubsection{Combined orienting and aligning interactions} 

\begin{turnpage}
\begin{figure}[H]
	\begin{subfigure}{.45\textwidth}
		\includegraphics[scale=0.55]{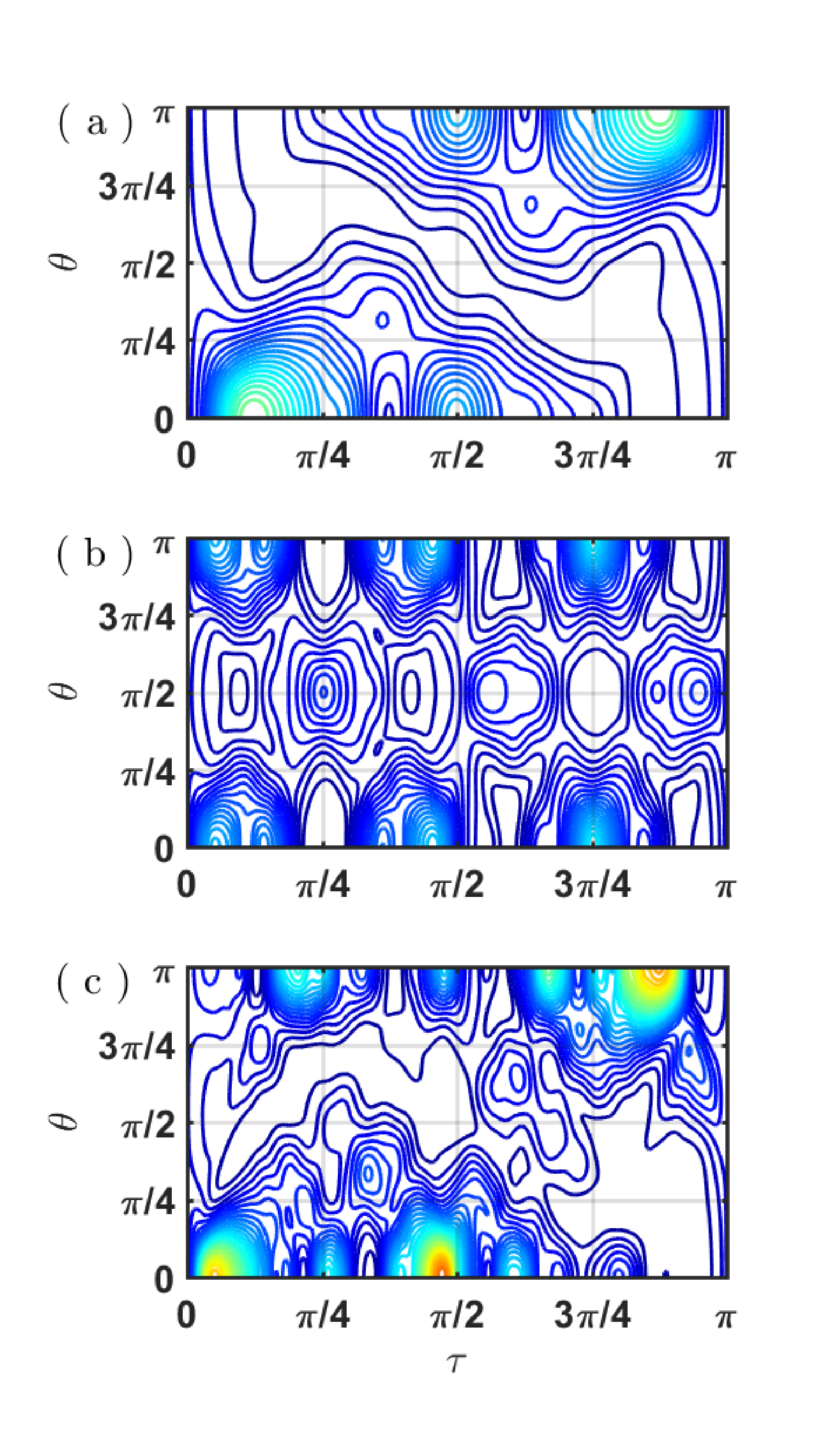}
	\end{subfigure}
	\begin{subfigure}{.45\textwidth}
		\includegraphics[scale=0.55]{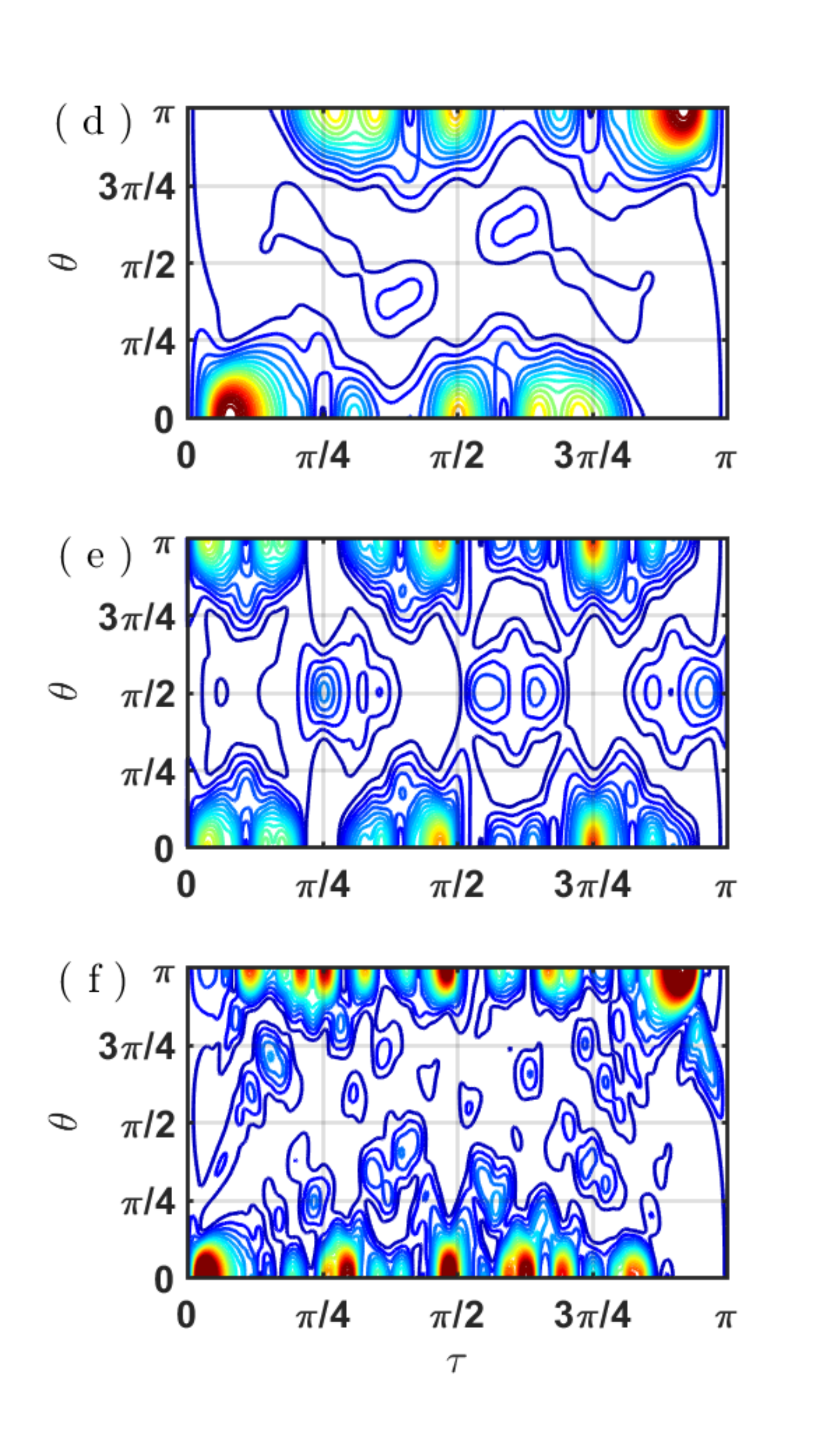}
	\end{subfigure} 
	\begin{subfigure}{.45\textwidth}
       \includegraphics[scale=0.55]{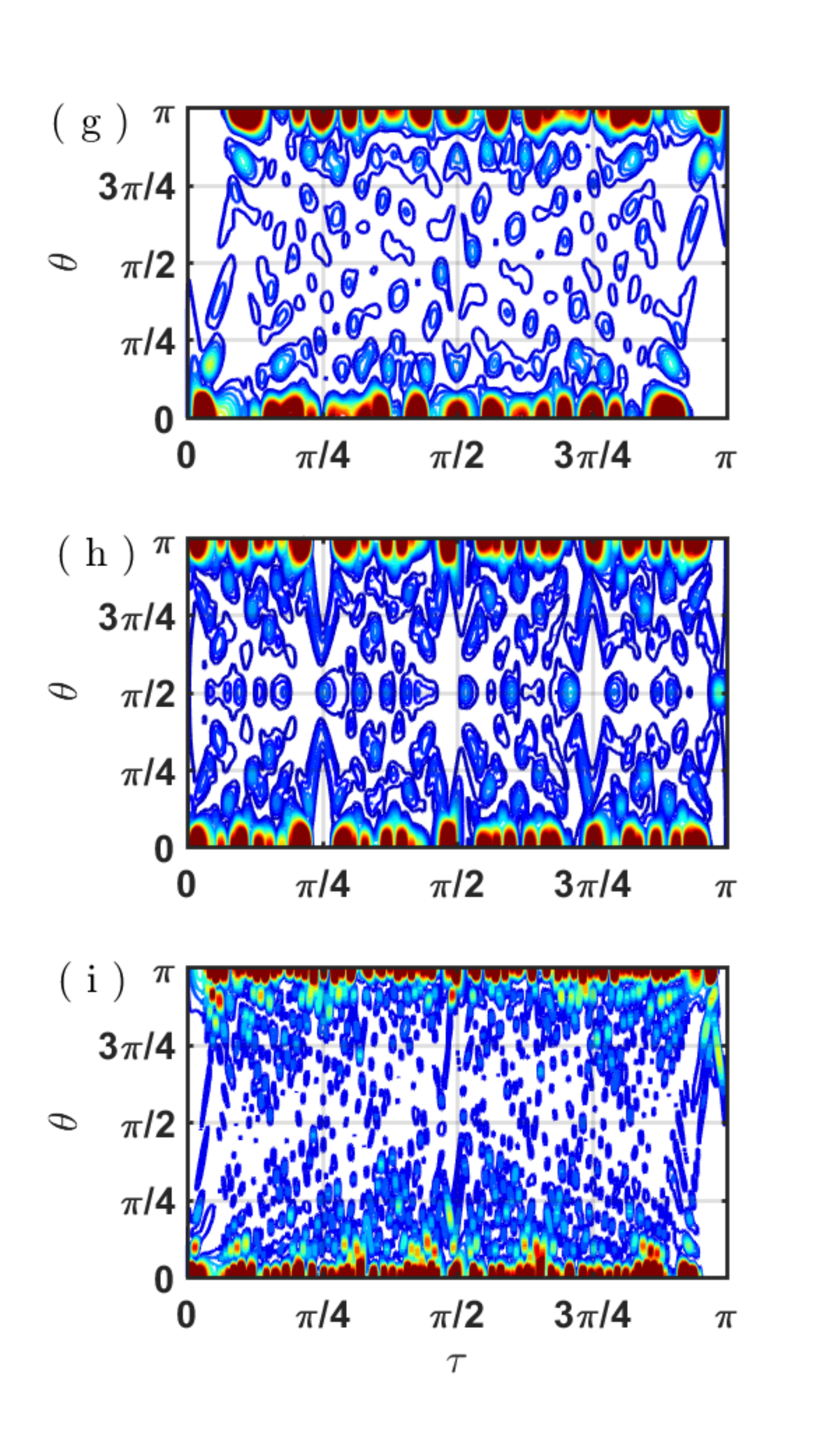}
    \end{subfigure}
	\caption{\label{carpets} Quantum carpets, $\left|\psi(\theta,\tau) \right|^2$ versus $\tau$, as obtained from Eq.~\eqref{prphasemodulus} for $\delta$-kicks with $P_\eta = P , P_\zeta= 0$ (panels a, d, g), $P_\zeta = P , P_\eta = 0$ (panels b, e, h), $P_\eta = P_\zeta = P$ (panels c, f, i), with $P = 1.5$ (panels a-c), $P = 2.8$ (panels d-f), and $ P = 8 $ (panels g-i).}
\end{figure}
\end{turnpage}

Panels (c), (f), and (i) of Fig.~\ref{carpets} show the quantum carpets for the combined orienting and aligning interactions of equal kick strength, $P_{\eta} = P_{\zeta}\equiv P$, from low ($P=1.5$), to intermediate ($P=2.8$), to high ($P=8$). 
For higher kick strengths, the number of fractional revivals beyond those discussed above (at full, half, and quarter of the rotational period) increases dramatically. As a result, the probability density patterns become complex and so does the analysis of Eq.~(\ref{prphasemodulus}) due to the presence of many terms. However, one can single out the following qualitative features, such as the ``wavepacket focusing'' and the formation of ``canals and ridges.'' 

\paragraph{Wavepacket focusing} was first observed for the purely orienting kicks and consists in the localization of the initially isotropic wavepacket around $\theta=0$ after short times $\tau_{\mathrm{f}}\approx1/(2P_{\eta})$ (this expression becomes exact at high $P_{\eta}$). By invoking the definition of wavepacket focusing based on the same semi-classical arguments as in Refs.~\cite{Averbukh2002,Leibscher2004}, we estimate the focusing time for the combined orienting and aligning interactions as $1/(2P_{\eta}+4P_{\zeta})$. 

We also encounter the phenomenon of  ``reversed-focusing'', i.e., localization of the wavepacket around $\theta = \pi$ shortly before the full revival. While for the purely orienting interaction, reversed-focusing occurs at $\tau_{\mathrm{rf}}=\tau_{\mathrm{rev}}-\tau_{\mathrm{f}}$, it comes about somewhat earlier for the combined interaction. 

\begin{figure}[H]
	\centering
	\includegraphics[scale=0.43]{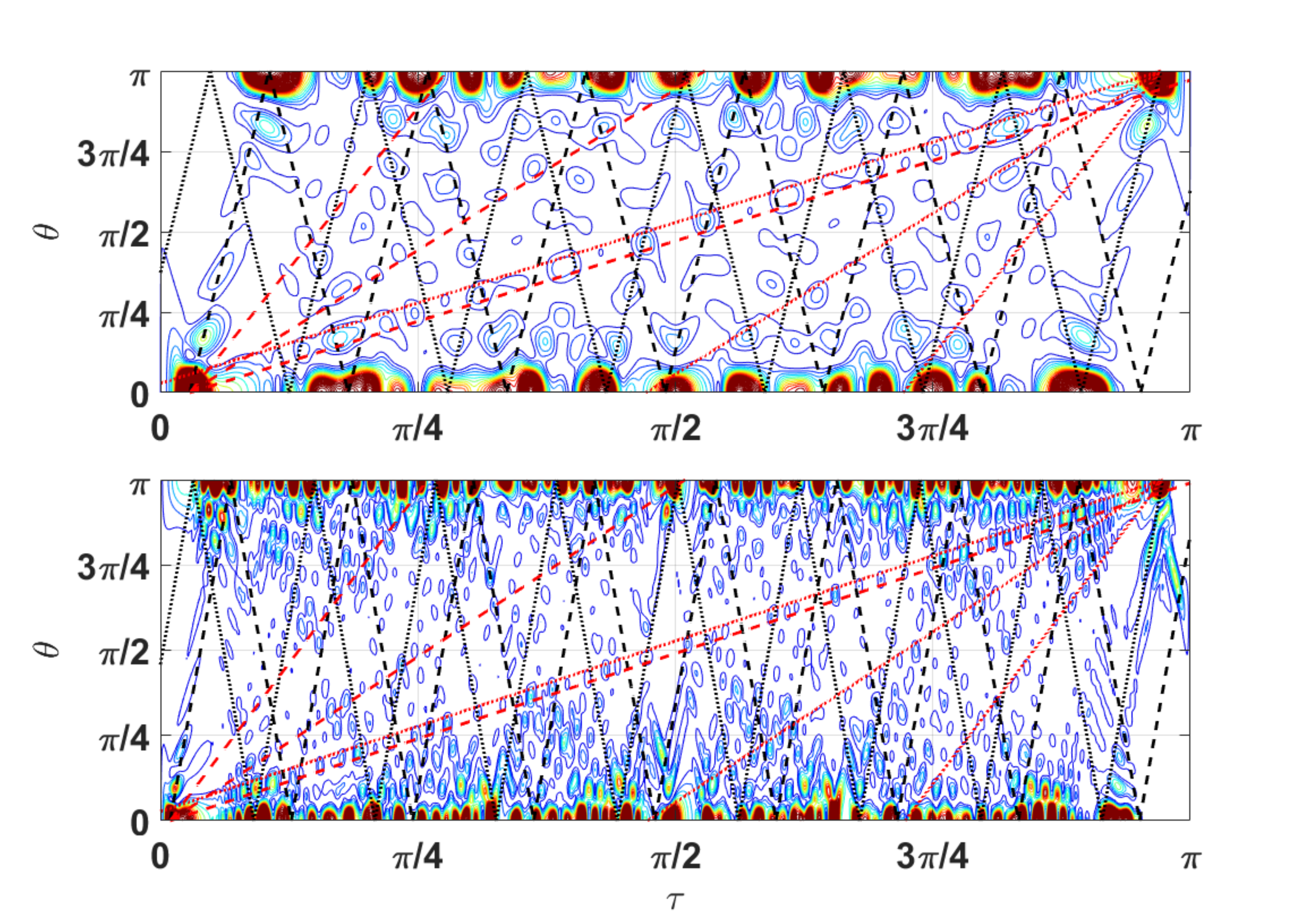}
	\caption{\label{carpets_lines} Same as panels (g) and (i) of Fig.~\ref{carpets} with black dashed lines showing the classical trajectories of Eqs.~\eqref{cr1},~\eqref{cr2} with $0\leq\beta\leq 12$ for upper panel and $0\leq\beta\leq 16$ for lower panel. Red dashed lines show the trajectories related to the fractional revivals of Eq.~\eqref{cr3} with $\nu\in\left\lbrace 1,\frac{1}{2},\frac{1}{4}\right\rbrace$. The dotted lines have the same meaning as the above ones but start from the reversed-focusing point.}
\end{figure}

\paragraph{Canals and ridges} are linear structures corresponding to low and high probability densities in the quantum carpets that arise at high kick strengths, see panels (g) and (i) of Fig.~\ref{carpets} as well as Fig.~\ref{carpets_lines}.

As shown in Refs. \cite{Grossmann1997,Harter2001,Robinett2004} these rich interference patterns are correlated with the space-time structure of the full and fractional revivals. Moreover, the ``characteristic rays'' indicating the loci of the canals and ridges can be interpreted as space-time trajectories of free particles on a ring propagating with a quantized angular velocity $\mathrm{d}\theta/\mathrm{d}\tau$ (in units of $B/\hbar$) \cite{Grossmann1997,Born1958a,Leibscher2009}. We note that the patterns of the carpets in Fig.~\ref{carpets} for $P_{\zeta}=0$ are similar to those reported in Ref. \cite{Leibscher2009} for the space-time evolution of a wavepacket corresponding to a squeezed pendular state initially localized in one of the potential wells
of a symmetric double well potential (with $\theta \in \left[ 0,2\pi\right]$).

The kinematic equations for the characteristic rays -- their ``classical background'' -- can be derived from  the average angular momentum  $\bar{J}$ of the system defined by \cite{Rozmej1998}, 
\begin{align}
\bar{J}(\bar{J}+1)\equiv \sum_{J} J(J+1) \left|C_{0,0}^J\right|^2
\end{align}
Using Eqs.~\eqref{kinetic} and \eqref{kin_tauplus}, we obtain
\begin{align}\label{jmean}
\bar{J}(P_{\eta},P_{\zeta}) =  \frac{1}{2} \left( \sqrt{\frac{8}{3} P_{\eta}^2 + \frac{32}{15} P_{\zeta}^2 + 1 }  - 1 \right) .
\end{align}
Then the classical dimensionless time $\tau_{\mathrm{cl}}$ is given by \cite{Robinett2004,Rozmej1998,Styer2001}
\begin{equation}
\tau_{\mathrm{cl}} = \frac{\pi}{\frac{\mathrm{d}E_J}{\mathrm{d}J}|_{J=\bar{J}}} = \frac{\pi}{(2\bar{J}+1)} .
\end{equation}

For instance, for $P_{\eta}=8$ and $P_{\zeta}=0$,  Eq.~\eqref{jmean} yields $\bar{J}(P_{\eta}=8 , P_{\zeta}=0)=6.0511$;  and for $P_{\zeta}=P_{\eta}=8$, $\bar{J}(P_{\eta}=8 , P_{\zeta}=8)= 8.2778$. Rounding these values to ($\left[ \bar{J}(8,0)\right]=6$ , $\left[ \bar{J}(8,8)\right]=8$) in order to fulfill the boundary condition for $\tau\in[0,\pi]$ and choosing $\tau_{\mathrm{f}}$ as the initial point, we obtain the classical trajectories \cite{Born1958,Born1958a} 
\begin{align}\label{cr1}
\theta(\tau) = \frac{\pi}{\tau_{\mathrm{cl}}} (\tau-\tau_{\mathrm{f}}) - \beta\pi
\end{align}
with $ \beta$  an even integer for rays running between $\theta=0$ to $\theta=\pi$ and
\begin{align}\label{cr2}
\theta(\tau) = -\frac{\pi}{\tau_{\mathrm{cl}}} (\tau-\tau_{\mathrm{f}}) + (\beta+1)\pi
\end{align}
with $ \beta$ an odd integer for rays running between $\theta=\pi$ to $\theta=0$. Note that $\beta$ indicates the number of reflections from the $\theta =0 $ and $\theta = \pi$ boundaries. 
We also note that taking $\tau_{\mathrm{f}}$ as the initial time is consistent with the localization of the initial wavepacket as required by the correspondence principle \cite{cohen1977,Averbukh1989a}. 

The trajectories given by Eqs.~\eqref{cr1} and \eqref{cr2} are shown by black dashed lines in Fig.~\ref{carpets_lines}. The dotted lines in Fig.~\ref{carpets_lines} start from a reversed-focused wavepacket at $\tau_{\mathrm{rf}}$ before the full revival time and propagate with the same slope as given by Eqs.~\eqref{cr1} and \eqref{cr2} but with decreasing $\tau$. These rays are of interest because of the backward bouncing in $\tau$.

In addition, some of the loci (without reflections) due to the fractional revivals are shown as red dashed and dotted lines in Fig.~\ref{carpets_lines}. The dashed red rays emerge from the focusing point close to the lower left corner of the space-time density plots ($\theta = 0, \tau \approx \tau_{\mathrm{f}}$) with the shape
\begin{align}\label{cr3}
\theta(\tau) = \frac{\pi}{\nu\tau_{\mathrm{rev}}} (\tau-\tau_{\mathrm{f}}) ,
\end{align}
where $\nu$ is a fraction of mutually prime integers (see also Refs. \cite{Averbukh1989a,Averbukh1989,Rozmej1998} for an extensive discussion of these rays and the fractional revivals). The dotted lines starting from the reversed-focusing point near the upper right corner ($\theta = \pi , \tau \approx \tau_{\mathrm{rf}}$) can be expressed as 
\begin{align}\label{cr4}
\theta(\tau) = \frac{\pi}{\nu\tau_{\mathrm{rev}}} (\tau-\tau_{\mathrm{rf}}) + \pi .
\end{align}

We note that it is the interference between the loci of the fractional revivals, as given by Eqs.~\eqref{cr3},~\eqref{cr4}, with the classical trajectories, as given by Eqs.~\eqref{cr1} and \eqref{cr2}, that gives rise to the rich pattern of the canals and ridges seen in Fig.~\ref{carpets_lines}.

\subsection{Field-free orientation and alignment due to $\delta$-kicks}\label{oriali}

We now investigate the field-free evolution of the orientation and alignment of a polar and polarizable rotor due to $\delta$-kicks within the analytic theory developed in Section \ref{Non-adia} for the sudden limit. We characterize the orientation and alignment of the rotor by the expectation values of the orientation and alignment cosines, $\left\langle \cos\theta  \right\rangle(\tau)$ and $\left\langle \cos^2\theta  \right\rangle(\tau)$, respectively.
We thus evaluate integrals of the form $\langle \psi(\theta,\tau) | \cos^{\lambda} \theta | \psi(\theta,\tau) \rangle$ with $\lambda = 1,2$ \cite{Arfken2005}. For an initial state $Y_0^0$, we obtain from Eqs.~\eqref{psi_final}, \eqref{psi_kicked_00}, and \eqref{C_lm0},

\begin{align}\label{orientation}
\left\langle \cos\theta  \right\rangle (\tau) = \sum_{J} \left\{ \frac{\mathrm{e}^{2i(J+1)\tau} (J+1)}{\sqrt{(2J+1)(2J+3)}}\left(C_{0,0}^{J+1}\right)^* C_{0,0}^J + \frac{ \mathrm{e}^{-2iJ\tau}J}{\sqrt{(2J-1)(2J+1)}} \left(C_{0,0}^{J-1}\right)^* C_{0,0}^J \right\}
\end{align}
and
\begin{align}\label{alignment}
\left\langle \cos^2\theta  \right\rangle (\tau) &= \sum_{J} \left\{ \left(\frac{1}{3}+\frac{2J(J+1)}{3(2J-1)(2J+3)}\right)\left|C_{0,0}^J\right|^2  + \frac{\mathrm{e}^{2i(2J+3)\tau}(J+1)(J+2)}{(2J+3)\sqrt{(2J+1)(2J+5)}}\left(C_{0,0}^{J+2}\right)^* C_{0,0}^J \right . \nonumber \\
&+ \left . \frac{\mathrm{e}^{-2i(2J-1)\tau}J(J-1)}{(2J-1)\sqrt{(2J-3)(2J+1)}} \left(C_{0,0}^{J-2}\right)^* C_{0,0}^J \right\} \nonumber \\
& = \left\langle \cos^2\theta  \right\rangle_p + \left\langle \cos^2\theta  \right\rangle_c (\tau)
\end{align}
where $C_{0,0}^{-2} = C_{0,0}^{-1} = 0$.
Here, $\left\langle \cos^2\theta  \right\rangle_p$ refers to the time-independent term $ \propto\left|C_{0,0}^J\right|^2$, in analogy to the population part of the probability density, the $\left\langle \cos^2\theta  \right\rangle_c$ term incorporates the time-dependence of the alignment due to the coherences. Whereas orientation involves $J$-state mixing (hybridization) that goes by the selection rule $\Delta J = \pm 1$,  for alignment the hybridization of $J$ obeys $\Delta J = \pm 2 , 0$.

Like the  wavefunctions, also orientation and alignment recur with a period of $\tau_{\mathrm{rev}}=\pi$, 
\begin{align}
\left\langle \cos^2\theta  \right\rangle (\tau+\pi) &=\left\langle \cos^2\theta  \right\rangle (\tau) \nonumber \\
\left\langle \cos\theta  \right\rangle (\tau+\pi) &=\left\langle \cos\theta  \right\rangle (\tau) 
\end{align}    
Since Eq.~\eqref{orientation} is a Fourier sum over terms with periods $\tau_{\mathrm{rev}}/(J+1)$ and $\tau_{\mathrm{rev}}/J$ can be even or odd integer fractions of the revival time.
Thus depending on the populations of different rotational states at the end of the pulse, $\left\langle \cos\theta  \right\rangle (\tau)$ can cycle several times before the full revival time is reached. 

Using Eq.~\eqref{orientation}, it is possible to show that the time-averaged orientation vanishes over a full revival interval for all kick strengths:
\begin{align}
\int\limits_{\tau_0\approx 0}^{\tau_{\mathrm{rev}}}\left\langle \cos\theta \right\rangle(\tau) \mathrm{d}\tau = 0
\end{align}

In contrast, the alignment, as given by Eq.~\eqref{alignment}, is a Fourier sum over terms with only odd integer fractions of the revival time, $\tau_{\mathrm{rev}}/(2J-1)$ and $\tau_{\mathrm{rev}}/(2J+3)$. Therefore, $\left\langle \cos^2\theta  \right\rangle (\tau)$ oscillates as a $\pi/2$-antiperiodic function with respect to the constant population term of Eq.~\eqref{alignment}, 
\begin{align}
\left\langle \cos^2\theta  \right\rangle_c \left(\tau+\frac{\pi}{2}\right) = -\left\langle \cos^2\theta  \right\rangle_c (\tau) 
\end{align}

This means that a reflected revival of coherent alignment (same as alignment of the initial state but with a changed sign) occurs at $\tau_{\mathrm{rev}}/2 $ in all cases,\begin{align}
\left\langle \cos^2\theta  \right\rangle_c \left(\frac{\pi}{2}\right) = -\left\langle \cos^2\theta  \right\rangle_c (0)   
\end{align}
see Figures~\ref{oriali15}-\ref{oriali8}. The left panels of these figures show the Fourier transforms (power spectra) of the (time-dependent) orientation and alignment signals. They display peaks at frequencies corresponding to the differences between the energies of free rotor states that are significantly populated by the pulse. The spectra of the orientation, \eqref{orientation}, show peaks located at Bohr frequencies $0,2,4,\ldots$ compatible with the selection rule $\Delta J=\pm 1$, whereas the peaks in the spectra of alignment, \eqref{alignment}, appear at $0, 6, 10, \ldots$, in accordance with the selection rule $\Delta J=\pm 2, 0$.

We now look at the behaviour of the field-free orientation and alignment as it arises due to the individual orienting and aligning interactions as well as their combination.

\subsubsection{Purely orienting interaction}
 
\begin{figure}[H]
	\centering
		\includegraphics[scale=0.32]{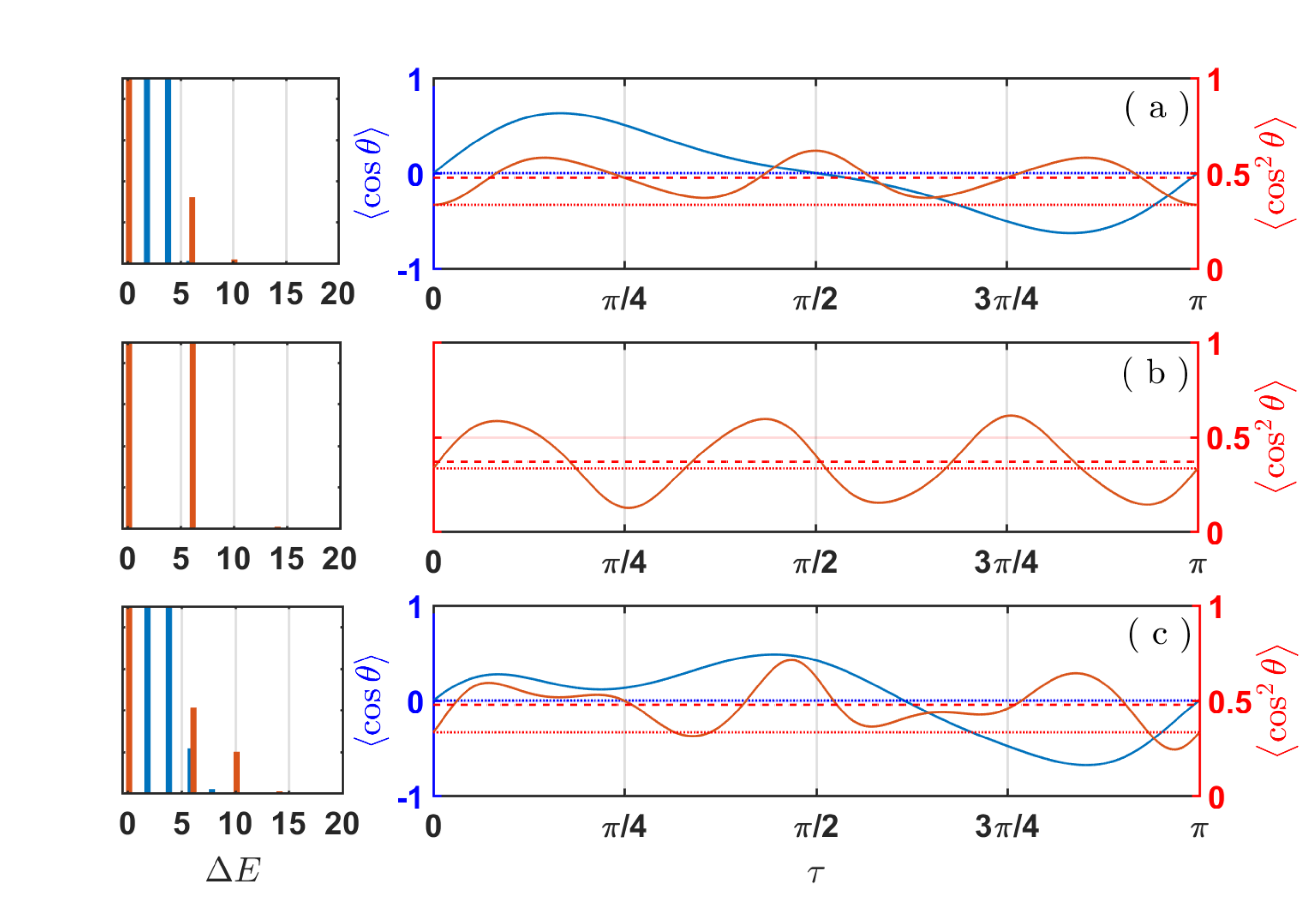}
		\caption{\label{oriali15}Plots of field-free evolution of the orientation cosine, $\left\langle \cos\theta \right\rangle(\tau) $, as given by Eq.~\eqref{orientation}, and alignment cosine, $\left\langle \cos^2\theta \right\rangle(\tau) $, as given by Eq.~\eqref{alignment}), due to an $\delta$-kick of strength (a)  $P_\eta = 1.5 , P_\zeta= 0$, (b) $P_\zeta = 1.5 , P_\eta = 0 $ and (c) $P_\eta = P_\zeta = 1.5 $. The blue dotted line marks the zero orientation cosine (an unoriented distribution) and the red dotted line an alignment cosine of $1/3$ (corresponding to an anisotropic spatial distribution). The red dashed horizontal line shows $ \left\langle \cos^2\theta \right\rangle_p $. Note that $\tau=0$ corresponds to $\tau_0$. Left panels show the Fourier transforms (power spectra) at frequencies $\Delta E$ that correspond to the differences between the energies of the free rotor states that are significantly populated  by the pulse. }
\end{figure}

\begin{figure}[H]
	\centering
		\includegraphics[scale=0.32]{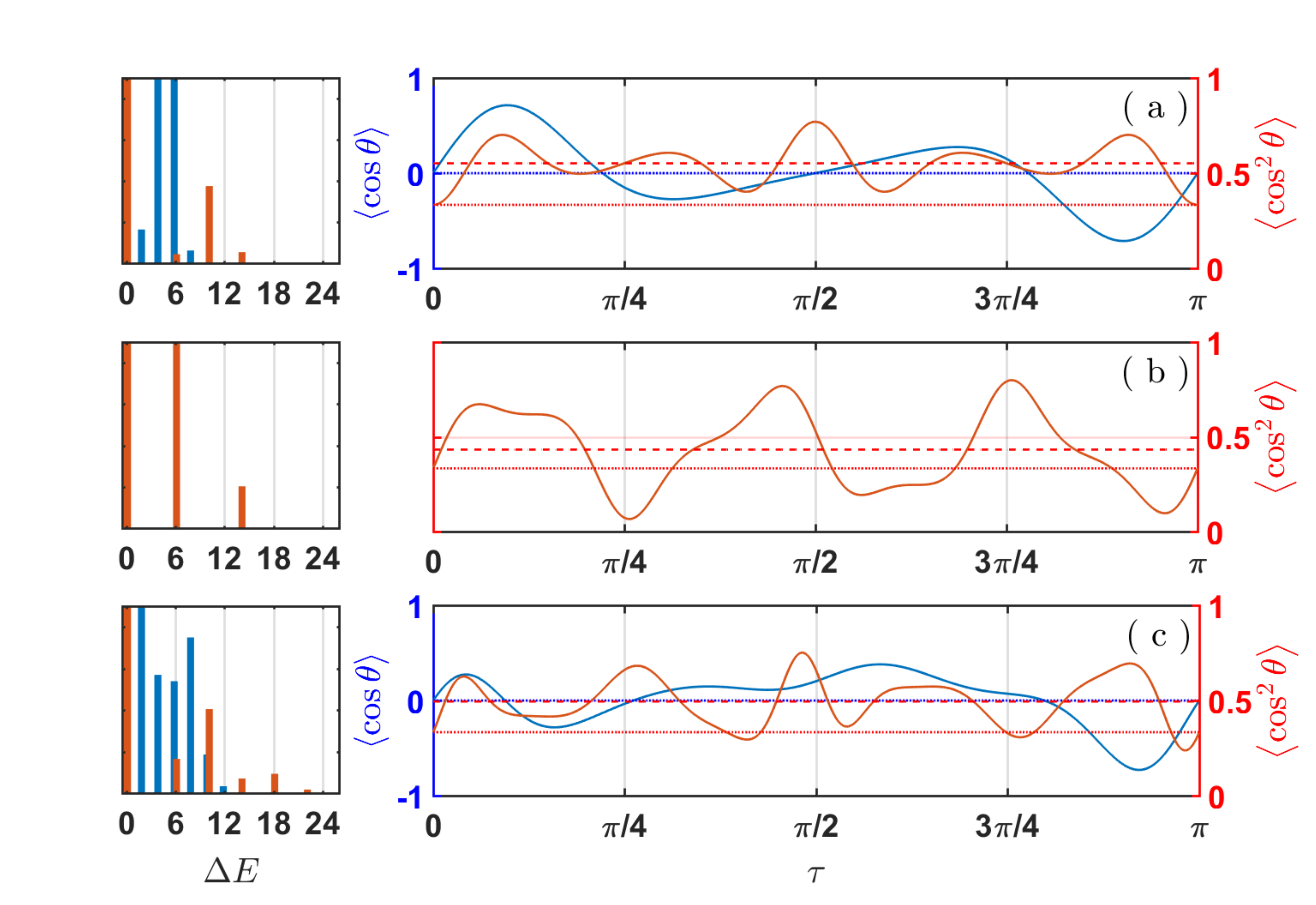}
		\caption{\label{oriali28} Same as Fig.~\ref{oriali15} with kick strengths (a) $P_\eta = 2.8 , P_\zeta= 0$, (b) $P_\zeta = 2.8 , P_\eta = 0 $ and (c) $P_\eta = P_\zeta = 2.8 $.}
\end{figure}

\begin{figure}[H]
	\centering
		\includegraphics[scale=0.32]{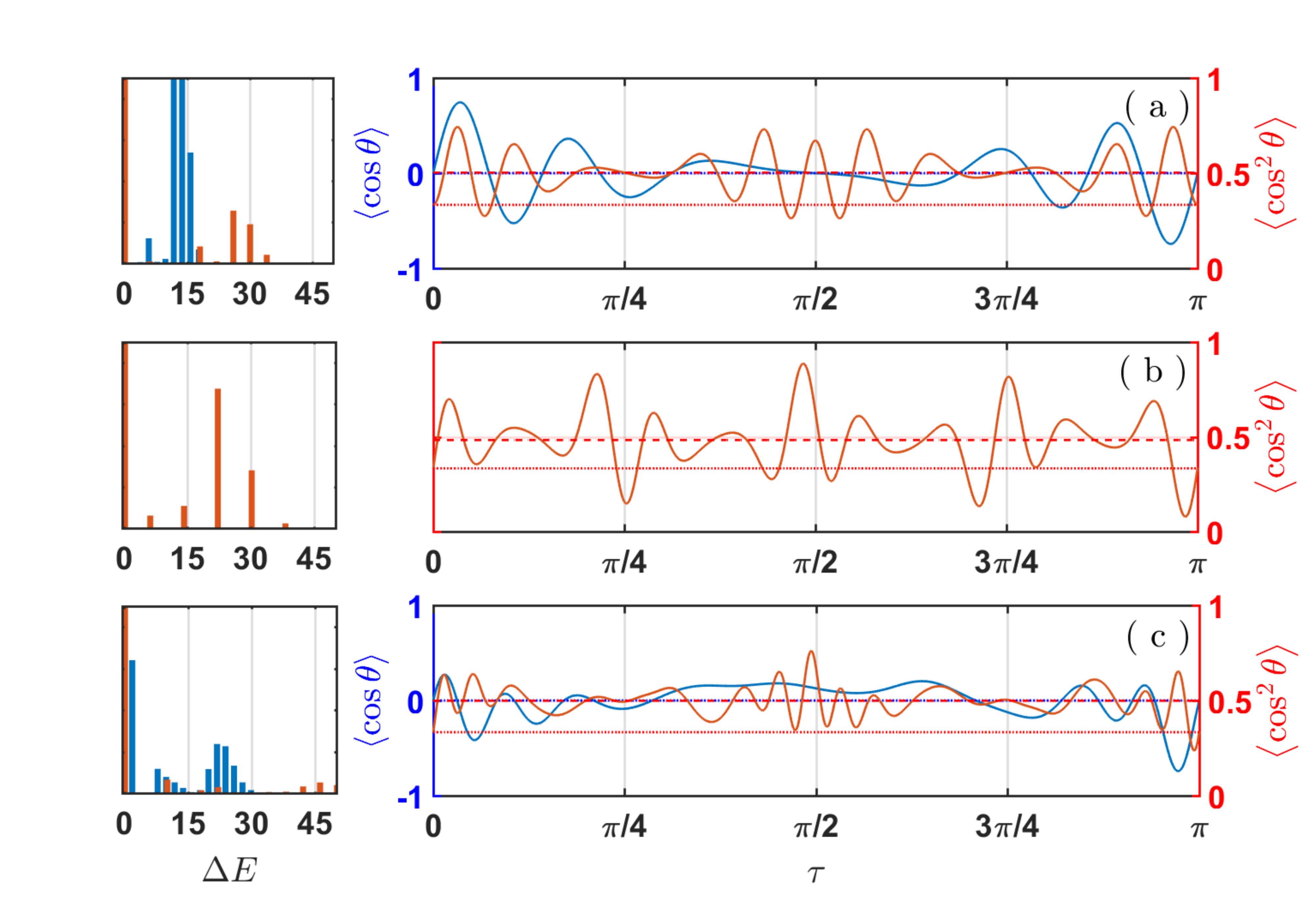}
		\caption{\label{oriali8} Same as Fig.~\ref{oriali15} with kick strengths (a)  $P_\eta = 8 , P_\zeta= 0$, (b) $P_\zeta = 8 , P_\eta = 0 $ and (c) $P_\eta = P_\zeta = 8 $.}
\end{figure}

By substituting for the coefficients $C_{0,0}^J$ from Eq.~\eqref{finalcoeff_00_pzeta0} into Eq.~\eqref{orientation} yields  
\begin{align}\label{oriphasemodul}
\left\langle \cos\theta  \right\rangle (\tau) &=  \sum_{J} \left\{ \frac{(J+1)}{\sqrt{(2J+1)(2J+3)}}\left| C_{0,0}^{J+1}\right| \left|  C_{0,0}^J\right| \cos\left(2(J+1)\tau-\frac{\pi}{2}\right) \right . \nonumber \\
& + \left . \frac{J}{\sqrt{(2J-1)(2J+1)}} \left| C_{0,0}^{J-1}\right| \left| C_{0,0}^J \right| \cos\left(-2J\tau+\frac{\pi}{2} \right) \right\}
\end{align}
where we made use of the identity $\Delta\gamma_{JJ\pm1} = \mp \pi/2$. Therefore, $\left\langle \cos\theta  \right\rangle (\tau) $ is an odd function with respect to the half of revival time,
\begin{align}
\left\langle \cos\theta  \right\rangle \left(\frac{\pi}{2}-\tau\right) = -\left\langle \cos\theta  \right\rangle \left(\frac{\pi}{2}+\tau\right)   
\end{align}
as illustrated by the blue curves in panels (a) of Figures~\ref{oriali15} and \ref{oriali8}.

Note that while the orientation at the end of the pulse, $\left\langle \cos\theta  \right\rangle (\tau_0)$, vanishes due to the spherical angular distribution (for which the alignment $\left\langle \cos^2\theta  \right\rangle (\tau_0) = \frac{1}{3}$), at the half-revival time, $\left\langle \cos\theta  \right\rangle (\frac{\pi}{2})$, the orientation vanishes because of the  alignment along the $z$ axis and therefore coincides with maximum alignment, see also panels (a), (d), and (g) of Fig.~\ref{carpets}. 

Inserting $\Delta\gamma_{JJ\pm2} = \mp \pi$ for the purely orienting interaction into Eq.~\eqref{alignment} gives
\begin{align}\label{alignphasemodul}
\left\langle \cos^2\theta  \right\rangle_c (\tau) &=  \sum_{J} \left\{  \frac{(J+1)(J+2)}{(2J+3)\sqrt{(2J+1)(2J+5)}}\left| C_{0,0}^{J+2}\right| \left|  C_{0,0}^J\right| \cos\left(2(2J+3)\tau-\pi\right) \right . \nonumber \\
& + \left . \frac{J(J-1)}{(2J-1)\sqrt{(2J-3)(2J+1)}} \left| C_{0,0}^{J-2}\right| \left| C_{0,0}^J \right| \cos\left(-2(2J-1)\tau+\pi \right) \right\} 
\end{align}
It is now easy to show that the coherence part of the alignment is an even function with respect to $\tau=\tau_{\mathrm{rev}}/2$,
\begin{align}
\left\langle \cos^2\theta  \right\rangle_c \left(\frac{\pi}{2}+\tau\right) = \left\langle \cos^2\theta  \right\rangle_c \left(\frac{\pi}{2}-\tau\right)
\end{align} 
and also 
\begin{align}
\left\langle \cos^2\theta  \right\rangle_c \left(\frac{\pi}{4}\right) = \left\langle \cos^2\theta  \right\rangle_c \left(\frac{3\pi}{4}\right) =0
\end{align}
Furthermore, $\left\langle \cos^2\theta  \right\rangle_c(\tau)$ is an odd function 
with respect to $\tau=\pi/4$ and $\tau=3\pi/4$, i.e., in the intervals $\left[ 0,\pi/2\right] $ and $\left[ \pi/2,\pi\right] $. It is shown by the red curves in panels (a) of Fig.~\ref{oriali15}.

\subsubsection{Purely aligning interaction} 
As shown before, for a purely aligning interaction, the wavepacket (for $Y_{0}^{0}$ as the initial state) is a superposition of spherical harmonics with even values  of quantum number $J$. Therefore it is not possible to fulfil the  selection rule for orientation, $\Delta J= \pm 1$,  and, consequently, $ \left\langle \cos\theta \right\rangle (\tau) $ vanishes. 
The  anti-alignment at $\tau\approx \pi/4$ and alignment at $\tau\approx 3\pi/4$, expected from the wavefunction analysis in Section~\ref{S-T_Pzeta}, is indeed confirmed in panels (b) of Figures~\ref{oriali15} and \ref{oriali8}. Like in the case of the probability densities, there is a transition between alignment and anti-alignment before and after the half revival time, respectively, as well as at the full revival time, albeit in an reversed order. The duration of the transition from alignment to anti-alignment depends on the kick strengths and in case of stronger kicks it gets dramatically reduced. For instance, for a weak kick of $P_{\eta}=0$ and $P_{\zeta}=1.5$, the transition occurs during $\tau_r/6$, whereas it is reduced to about $\tau_r/24$ for $P_{\eta}=0,P_{\zeta}=8$.

\subsubsection{Combined orienting and aligning interactions} 

In the case of the combined orienting and aligning interactions (of equal kick strengths), $\left\langle\cos\theta\right\rangle(\tau) > 0$ for over half a revival period, see panels (c) of Figs.~\ref{oriali15} to \ref{oriali8}.
However, the maximum absolute value of the orientation cosine corresponds to anti-orientation ($\left\langle\cos\theta\right\rangle_{max} < 0$), which lasts for a shorter time at higher kick strengths.

We also note a transition from aligned to nearly isotropic angular distribution, $\left\langle \cos^2\theta  \right\rangle \approx 1/3$, before and after the half-revival, respectively. The duration of this transition is reduced for stronger kick strengths, e.g, for  $P_{\eta}=P_{\zeta}=8$, it is about a fifth of what it is for $P_{\eta}=P_{\zeta}=1.5$. 

\section{Conclusions and prospects}\label{conclusions}

We have developed an analytic theory of the interaction of polar and polarizable rotors with plane-polarized unipolar $\delta$-pulses and used it to study the dynamics of this interaction systematically. We considered, in turn, the purely orienting, purely aligning, and combined orienting and aligning dynamics for a range of parameters on which the sudden-limit dynamics depends, namely the dimensionless orienting and aligning kick strengths. We found that the dynamics is interaction-specific, differing greatly for the orienting and aligning interactions as well as for their combination. 

The populations of the rotational states created by the $\delta$-kick as captured by the population quilts exhibit characteristic differences reflecting the fact that the orienting and aligning interactions obey different selection rules for the rotational quantum number $J$.  Interestingly, for the combined interactions of equal kick strength, the most populated states are those with $J=0$ and $J=1$. 

The space-time dependence -- as rendered by the quantum carpets -- of the recurring rotational wavepackets created by the $\delta$-kicks also exhibits sui generis behaviour for the orienting, aligning, and combined interactions. Even though the full revival time is independent of the type of interaction, the quantum carpets have conspicuously different patterns as fractional revivals occur with distinct interaction- and kick strength-dependent probabilities. Also the maximum values of the orientation and alignment achieved differ markedly for the three types of interaction and recur with different transition times. Some of the dynamics manifest in the quantum carpets we were able to explain qualitatively by invoking symmetry (parity) and the correspondence principle (canals and ridges). 

In addition, we studied the effects of finite-width pulses all the way up to the well-understood adiabatic limit. These numerical calculations served to test our analytic results as well as to explore resonances in the angular momentum/kinetic energy imparted to the rotor as a function of the pulse duration. These resonances occur due to a sudden drop of the wave function coefficients pertaining to particular $J$ states as a function of pulse duration and are recognized as an essential  feature of the sudden-to-adiabatic transition regime. While an analytic expression for the positions of these resonances as a function of kick strength remains elusive (owing to divergent integrals involved), we have been able to unravel the general behaviour by means of a numerical analysis. We hope to expand our understanding of the dynamic resonances and to further explore their possible experimental signature as well as their use in applications, such as spectroscopy, field-free stereo-dynamics of molecular collisions, photodissociation, orbital imaging and others.

Although our analytic treatment-based calculations pertain to rotors that are initially in the $J=0$ ground state, our analytic theory is general enough to account for any other initial free-rotor state. In our future work, we will explore the effects of these other rotor states on the outcome of the three types of $\delta$-kicks. Another avenue of future research will be the exploration of any regularities (such as pulse duration, combination of orienting and aligning kick strengths) that govern the maximum achievable values of the orientation and alignment. 

\begin{acknowledgments}
	We dedicate this paper to Helmut Schwarz on the occasion of his 75th birthday, with admiration for his leadership in science and beyond.
	
	Support by the \textit{Deutsche Forschungsgemeinschaft} (DFG) through grants SCHM 1202/3-1 and FR 3319/3-1 is gratefully acknowledged.
\end{acknowledgments}

\appendix
\section{Evaluation of the $C_{J_0,M_0}^J$ coefficients}\label{app_CJ}

The coefficients  $C_{J_0,M_0}^J$ arising in Eq.~\eqref{wf_expan_har_2} can be found by expanding the time-independent term in Eq.~\eqref{time_eval_psi} in terms of spherical harmonics, 
\begin{gather}
	\mathrm{e}^{i\left(P_{\eta}\cos\theta + P_{\zeta}\cos^2\theta\right)} Y_{J_0}^{M_0}(\theta,\phi) = \sum_{J} C_{J_0,M_0}^J Y_{J}^{M_0} (\theta,\phi) ~, \nonumber \\
	C_{J_0,M_0}^J = \int_0^{2\pi}\mathrm{d}\phi \int_0^{\pi} \sin\theta \mathrm{d}\theta \left[Y_{J}^{M_0}(\theta,\phi)\right]^*\mathrm{e}^{i\left(P_{\eta}\cos\theta + P_{\zeta}\cos^2\theta\right)} Y_{J_0}^{M_0}(\theta,\phi) ~. \label{exp_yl0m0expan}
\end{gather}  
The exponential part of Eq.~\eqref{exp_yl0m0expan} is only a function of $\theta$ and not $\phi$ and so can be expanded in terms of Legendre polynomials $\mathcal{P}_{J}$, 
\begin{gather}
	\mathrm{e}^{i\left(P_{\eta}\cos\theta + P_{\zeta}\cos^2\theta\right)} = \sum_{J^\prime} c^{J'} \mathcal{P}_{J^\prime}(\cos\theta) ~, \nonumber \\
	c^{J'} = \frac{2J^\prime+1}{2} \int_0^{\pi} \mathcal{P}_{J^\prime}(\cos\theta)  \mathrm{e}^{i\left(P_{\eta}\cos\theta + P_{\zeta}\cos^2\theta\right)} \sin\theta \mathrm{d}\theta \label{exp_expan}
\end{gather}  
where the factor $\frac{2J^\prime+1}{2}$ arises due to the orthogonality of the Legendre polynomials. With the change of variable $ x \equiv \cos\theta$, we have
\begin{eqnarray}
	c^{J'} &=& \frac{2J^\prime+1}{2} \int_{-1}^{1} \mathcal{P}_{J^\prime}(x)  \mathrm{e}^{i\left(P_{\eta}x+ P_{\zeta}x^2\right)} \mathrm{d}x \label{b_l_1} \nonumber \\
	&=& \frac{2J^\prime+1}{2} \sum_{k=0}^\infty\sum_{\ell=0}^k \frac{i^k}{k!}\binom{k}{\ell} P_\eta^{k-\ell} P_\zeta^\ell \int_{-1}^1 x^{k+\ell} \mathcal{P}_{J^\prime}(x)\mathrm{d}x \label{b_l_2} ~.
\end{eqnarray}
In order to obtain the second line in Eq.~\eqref{b_l_1}, we make use of a series expansion of the exponential function and the binomial theorem. When  $k+\ell < J^\prime $ or $ k+\ell+J^\prime$ is an odd integer, the integral in Eq.~\eqref{b_l_2} will vanish and, as a result, $c^{J'} = 0$. Otherwise \cite{Gradshtein2000},
\begin{align}\label{integral}
	\int_{-1}^1 x^{k+\ell} \mathcal{P}_{J^\prime}(x)\mathrm{d}x &= \frac{\sqrt{\pi}}{2^{k+\ell}}\frac{\Gamma(k+\ell+1)}{\Gamma\left((k+\ell-J^\prime)/2+1\right) \Gamma\left((k+\ell+J^\prime)/2+3/2\right)} \nonumber \\
	&= \frac{1}{2^{J^\prime}} \binom{k+\ell}{J^\prime}\frac{\Gamma\left(\left(k+\ell-J'+1\right)/2\right) \Gamma\left(J'+1\right)}{\Gamma\left(\left(k+\ell+J^\prime+3\right)/2\right)}
\end{align}	
and
\begin{equation}\label{b_l}
	c^{J'} = \frac{2J^\prime+1}{2} \sum_{k=0}^\infty\sum_{\ell=0}^k \frac{i^k}{k!}\binom{k}{\ell} P_\eta^{k-\ell} P_\zeta^\ell \frac{1}{2^{J^\prime}} \binom{k+\ell}{J^\prime}\frac{\Gamma\left(\left(k+\ell-J'+1\right)/2\right) \Gamma\left(J'+1\right)}{\Gamma\left(\left(k+\ell+J^\prime+3\right)/2\right)}
\end{equation} 
with $\Gamma$ the gamma function; we applied the Legendre duplication formula \cite{Arfken2005} to achieve the final form of Eq.~\eqref{integral}.

By making use of Eq.~\eqref{exp_expan} and Eq.~\eqref{b_l} we can evaluate$C_{J_0,M_0}^J$ in Eq.~\eqref{exp_yl0m0expan} as follows,
\begin{eqnarray}
	C_{J_0,M_0}^J &=& \int_0^{2\pi}\mathrm{d}\phi \int_0^{\pi} \sin\theta \mathrm{d}\theta \left[Y_{J}^{M_0}(\theta,\phi)\right]^*\left[\sum_{J^\prime=0}^\infty c^{J'} \mathcal{P}_{J^\prime}(\cos\theta)\right]Y_{J_0}^{M_0}(\theta,\phi) \nonumber \\
	&=& \sum_{J^\prime=0}^\infty c^{J'} \sqrt{\frac{4\pi}{2J^\prime+1}} \int_0^{2\pi}\mathrm{d}\phi \int_0^{\pi} \sin\theta \mathrm{d}\theta \left[Y_{J}^{M_0}(\theta,\phi)\right]^* Y_{J^\prime}^{0}(\theta,\phi) Y_{J_0}^{M_0}(\theta,\phi) \label{Alm_1} \nonumber \\
	&=& \sum_{J^\prime=0}^\infty c^{J'} \sqrt{\frac{2J_0+1}{2J+1}}  \left\langle J'0,J_00|J0 \right\rangle \left\langle J'0,J_0M_0|JM_0 \right\rangle \label{Alm_2}
\end{eqnarray} 
where $\left\langle J_1M_1, J_2M_2 |J_3M_3\right\rangle $ are the Clebsch-Gordan coefficients, which vanish unless $ |J - J_0| \leq J^\prime \leq J+J_0$ and $J+J^\prime+J_0$ is an even integer \cite{Arfken2005}.
Finally, we obtain 
\begin{equation}\label{C_lm0}
	C_{J_0,M_0}^J = \sum_{J^\prime=|J - J_0|}^{J+J_0} c^{J'} \sqrt{\frac{2J_0+1}{2J+1}} \left\langle J'0,J_00|J0 \right\rangle \left\langle J'0,J_0M_0|JM_0 \right\rangle .~
\end{equation}

In all the analytic results presented in this paper, Eqs.~\eqref{b_l} and \eqref{C_lm0} were found to converge satisfactorily for $\kappa$ and $J$ up to 80 and 50, respectively (e.g. for $P_{\eta}=P_{\zeta}=8$ the $80$th term of the sum over $k$ in Eq.~\eqref{b_l} is close to $10^{-30}$ for $ C_{0,0}^{50} \approx 10^{-15}$).

\newpage

\bibliography{HCP_combined_fields_ref} 
\end{document}